\documentclass[11pt]{article}

\usepackage[margin=1in]{geometry}
\usepackage[utf8]{inputenc}
\usepackage{times}
\usepackage{graphicx}
\usepackage{amsmath, amssymb}
\usepackage{caption}
\usepackage{subcaption}
\usepackage{booktabs}
\usepackage{multirow}
\usepackage{float}
\usepackage{placeins}
\usepackage{url}
\usepackage{listings}
\usepackage[most]{tcolorbox}
\usepackage{tikz}
\usepackage{abstract}
\usetikzlibrary{calc, positioning, arrows.meta, shapes.geometric}

\usepackage[colorlinks=true, linkcolor=black, citecolor=black, urlcolor=black]{hyperref}

\usepackage[sorting=none]{biblatex}
\graphicspath{{figures/}}
\lstdefinestyle{promptstyle}{
  basicstyle=\ttfamily\footnotesize,
  columns=fullflexible,
  breaklines=true,
  breakatwhitespace=true,
  keepspaces=true,
  showstringspaces=false
}

\title{\textbf{Design and Evaluation of Multi-Agent AI Oracle Systems for Prediction Market Resolution}}
\author{
Tarun Kota \\
Yale University \\
\texttt{Tkota0910@gmail.com}
}

\date{}

\begin{document}

\maketitle

\begin{abstract}
\thispagestyle{plain}

Prediction markets aggregate collective intelligence to forecast uncertain events, but their utility depends on reliable resolution of outcomes. Current oracle systems force a tradeoff between fast but brittle automation and accurate but costly human arbitration. Recent work has shown that single-LLM oracles can achieve meaningful accuracy, but these systems inherit all failure modes of their underlying model with no mechanism for self-correction. This thesis evaluates whether multi-agent LLM architectures can improve oracle resolution accuracy over single-model baselines. We compare two architectures, independent aggregation and deliberative consensus, against single-LLM baselines using three diverse models (GPT-5 Nano, DeepSeek V3, Llama-3.3-70B) on 1,189 resolved prediction market questions from KalshiBench. All agents share a common evidence layer through Exa, with retrieval filtered by publication date to isolate reasoning capability from retrieval quality. Independent aggregation with confidence-weighted voting achieves the highest accuracy at 83.43\%, outperforming the best individual model by 1.01 percentage points. Deliberative consensus degrades accuracy to approximately 76\%, below every single-model baseline, a result we attribute to persuasive error propagation, where confidently wrong models flip correct ones during debate. Moderate-to-high error correlations across models ($r = 0.529$--$0.689$) explain why aggregation gains fall well short of the theoretical Condorcet ceiling and place a fundamental limit on ensemble-based approaches. A substantial fraction of questions resist correction by any multi-agent architecture, motivating principled escalation to human arbitration. Based on these findings, we propose routing criteria for hybrid AI-human oracle systems: auto-resolving only unanimous, high-confidence questions yields 97.87\% accuracy on 47\% of the dataset, with inter-agent disagreement flagging the remainder for human arbitration.

\end{abstract}

\section{Introduction}
\subsection{Motivation and Context}

Prediction markets are financial instruments that allow participants to trade contracts whose payoffs depend on the outcome of real-world events, such as election results, economic indicators, or sporting outcomes. Because traders with superior information can profit by correcting mispriced contracts, market prices aggregate dispersed knowledge into probability estimates that have been shown to outperform polls, expert panels, and statistical models across a range of domains~\cite{arrow2008promise, berg2008prediction}. Platforms such as Polymarket, Kalshi, and Metaculus have grown rapidly in recent years. Polymarket processed over \$33 billion in trading volume in 2025, and Kalshi's revenue grew 994\% year-over-year after winning a landmark court case against the CFTC in September 2024~\cite{kalshi2024cftc}.

The utility of a prediction market, however, depends entirely on the reliability of its oracle, the mechanism that determines whether a real-world event actually occurred and settles contracts accordingly. This is an instance of the broader \emph{oracle problem} in blockchain systems: smart contracts cannot natively access off-chain information, so some trusted process must bridge the gap between on-chain logic and real-world state~\cite{caldarelli2020oracle, ellis2017chainlink}. Existing oracle designs occupy two extremes. Fully automated oracles such as Chainlink's decentralized oracle networks provide fast, inexpensive resolution for structured data feeds like asset prices, but struggle with subjective or ambiguous outcomes~\cite{chainlink2021v2}. Human arbitration systems such as UMA's optimistic oracle and Kalshi's internal operations team achieve higher accuracy on complex questions, but incur significant latency and escalation costs that scale poorly~\cite{uma2024oracle, kalshi_outcomes}. The scale of value at risk is substantial: DeFi protocols lost over \$400 million to oracle manipulation attacks in 2022 alone~\cite{caldarelli2020oracle}, and prediction market volumes now routinely exceed billions of dollars per month.

Recent work has demonstrated that large language models can serve as an intermediate resolution layer between full automation and human arbitration. Chainlink Labs validated a DSPy-based single-LLM oracle achieving 99.7\% accuracy on sports questions but only 84--85\% on politics and cryptocurrency~\cite{chainlink2025aioracles}, while UMA's OOTruthBot achieves approximately 90\% accuracy at roughly \$0.005 per resolution~\cite{uma2025truthbot}. These results are promising, but single-LLM systems inherit all failure modes of their underlying model, including hallucinations, temporal reasoning errors, and sycophantic tendencies, with no mechanism for self-correction. A natural extension is to employ multi-LLM architectures for oracle resolution. Yet this space remains largely unexplored.

\subsection{Research Question}

This paper addresses the oracle reliability gap through the following central research question:

\emph{Can multi-agent LLM architectures improve oracle resolution accuracy over single-model baselines for prediction markets, and under what conditions do such improvements occur?}

This question gives rise to several focused sub-questions that structure the evaluation:

\begin{enumerate}
    \item \textbf{Aggregation Effectiveness:} Does combining multiple LLMs into an independent-aggregation ensemble improve oracle accuracy relative to a single-model system?
    
    \item \textbf{Deliberation versus Independent Voting:} Does structured debate among LLMs outperform independent aggregation, or does it introduce conformity effects that degrade performance?
    
    \item \textbf{Failure Mode Correlation:} Are oracle-specific errors independent across models, or do they exhibit systematic correlation, and what distinct failure mechanisms produce the errors that aggregation fails to correct?
    
    \item \textbf{Escalation Design:} Can model confidence and inter-agent disagreement serve as principled routing signals for deferring uncertain questions to human arbitration?
\end{enumerate}

By answering these questions, this work provides empirical evidence on multi-agent oracle performance and develops a structured framework for hybrid AI-human resolution systems.

\subsection{Problem and Literature Gap}

Despite promising results from single-LLM oracle systems, no prior work has applied multi-agent techniques to prediction market resolution, where correctness directly impacts financial outcomes and system trust. Existing studies on multi-agent debate focus on general reasoning benchmarks, leaving unclear whether reported improvements transfer to the distinct challenges of oracle resolution: interpreting natural-language resolution criteria, verifying time-bounded real-world events, and adjudicating inherently ambiguous outcomes.

Moreover, no prior work has characterized error correlation specifically in oracle settings. While correlated failures across LLMs are documented in general reasoning tasks~\cite{kim2025correlated}, it remains unknown whether oracle-specific failure modes (source verification errors, misinterpretation of resolution criteria, temporal reasoning mistakes) exhibit similar or distinct correlation patterns. Without this understanding, aggregating multiple models may offer little robustness benefit and may instead amplify shared biases.

Finally, current hybrid AI-human oracle systems rely on ad-hoc escalation policies, with no principled framework for determining when model confidence is insufficient and human arbitration justifies its cost and latency. This lack of structure limits the scalability and reliability of hybrid oracle designs.

\subsection{Contributions}

This work makes three contributions:

\begin{itemize}
    \item We provide the first empirical comparison of multi-agent oracle architectures, independent aggregation and deliberative consensus, against single-LLM baselines for prediction market resolution, a domain where correctness has direct financial consequences.
    
    \item We develop a failure mode taxonomy that characterizes the distinct 
mechanisms by which the system fails, identifying five failure types and 
measuring the degree to which errors are shared across models versus 
architecture-specific.
    
    \item We propose escalation criteria for hybrid AI-human oracle systems, defining principled decision rules based on confidence levels and inter-agent disagreement for routing between automated and human resolution.
\end{itemize}

\subsection{Research Methodology}

We evaluate two multi-agent architectures using GPT-5-Nano, Deepseek V3, and Llama-3.3-70b-turbo:

\textbf{Architecture A: Independent Aggregation.} Three diverse LLMs resolve questions independently, with outcomes determined by majority voting and confidence-weighted voting. This mirrors how Chainlink aggregates price feeds from multiple node operators and tests whether the same logic applies to subjective outcomes.

\textbf{Architecture B: Deliberative Consensus.} The same three models participate in structured multi-round debate: each generates an answer with reasoning, then reviews the other models' responses and may revise its position before submitting a final answer.

Both architectures are evaluated against single-LLM baselines on KalshiBench, a benchmark of over 1,500 prediction market questions from Kalshi with verified real-world outcomes occurring after model training cutoffs~\cite{nel2025kalshibench}. Crucially, all agents share a common evidence layer through Exa, an embeddings-based search engine, with retrieval filtered by publication date to prevent training data contamination. This design isolates reasoning capability from retrieval capability, enabling controlled comparison of the architectures.

\subsection{Overview of Findings}

Independent aggregation with confidence-weighted voting achieves the highest overall accuracy at 83.43\%, a modest but significant improvement over the best single-model baseline (DeepSeek, 82.42\%). Deliberative consensus, contrary to expectations from the multi-agent debate literature, degrades performance to approximately 76\%, below every single-model baseline, a result we attribute to persuasive error propagation and LLM sycophancy, where confidently wrong models flip correct ones during debate. Error correlations across models are moderate to high ($r = 0.529$--$0.689$), explaining why majority-vote accuracy falls well short of the theoretical Condorcet ceiling and placing a fundamental limit on aggregation-based approaches. Category-level analysis reveals a clear tier structure, with well-documented domains like Sports and Financials exceeding 92\% while ambiguous domains like Crypto and Companies remain below 67\% across all architectures, and 14\% of questions resist correction by any multi-agent approach. These findings inform a principled escalation framework: unanimous, high-confidence resolutions achieve 97.87\% accuracy while covering nearly half of all questions, enabling hybrid oracle systems to 
auto-resolve straightforward cases while routing genuinely uncertain 
questions to human arbitration.

\section{Background}
This section provides the foundational context necessary to understand the design space for AI-based oracle systems. We introduce prediction markets, explain the oracle problem that constrains their reliability, survey existing resolution mechanisms, and motivate why large language models represent a natural intermediate solution.

\subsection{Prediction Markets and Information Aggregation}

A prediction market is a marketplace for contracts whose payoff depends on a future event. In a common binary design, a contract pays \$1 if the event occurs and \$0 otherwise. If such a contract trades at price $p \in [0, 1]$, practitioners often interpret $p$ as an implied probability, because a trader who believes the true probability is higher than $p$ has an incentive to buy, while a trader who believes it is lower has an incentive to sell or short. As new information arrives, profit-motivated traders update their positions, and the price moves to reflect the market's aggregated belief. At the market's end, the contract cash-settles based on an externally verified outcome, so correctness hinges on the integrity of the resolution process. This settlement step creates the oracle problem. The market can only function if a trusted mechanism can map real-world outcomes into an unambiguous on-chain or off-chain final result.

The intellectual foundation traces to Hayek's insight that market 
prices aggregate dispersed knowledge held by individuals throughout 
society~\cite{hayek1945use,arrow2008promise}. Empirically, prediction 
markets have been closer to eventual outcomes than polls 74\% of the 
time across five U.S.\ presidential elections~\cite{berg2008prediction}, 
and a meta-analysis of 160 publications concluded that prediction 
markets are 79\% more accurate than alternative forecast 
methods~\cite{iceb2020meta}.

\subsection{The Oracle Problem}

The oracle problem refers to the fundamental challenge of injecting reliable 
off-chain data into deterministic blockchain systems. Smart contracts cannot 
natively access external information because blockchain consensus requires 
all nodes to independently verify computations. External API calls would 
return different results at different times, breaking deterministic 
execution~\cite{ellis2017chainlink, caldarelli2020oracle}. Despite its 
centrality to blockchain applications, of 142 journal papers on blockchain 
applications, only 15\% considered the role of oracles, and fewer than 
10\% discussed oracle limitations~\cite{caldarelli2020oracle}.

The financial consequences of oracle failures are severe. DeFi protocols 
lost over \$400 million across 41 oracle manipulation attacks in 2022 
alone~\cite{chainalysis2023oracle}. No blockchain oracle protocol can 
simultaneously optimize scalability, decentralization, and truthfulness, 
a constraint formalized as the Oracle Trilemma~\cite{cong2025oracle}. 
This impossibility result motivates hybrid approaches that combine automated 
AI resolution with human oversight, rather than relying on either alone.

\subsection{Existing Oracle Resolution Systems}

Existing oracle systems can be understood along a spectrum from full automation to full human adjudication, with each design making different tradeoffs among speed, cost, accuracy, and trust assumptions.

\subsubsection{Chainlink's Decentralized Oracle Networks}

Chainlink introduced a two-layer architecture with on-chain aggregation of oracle node responses and off-chain data sourcing~\cite{ellis2017chainlink}. The Chainlink 2.0 whitepaper expanded this to Decentralized Oracle Networks (DONs), committees of nodes providing oracle services alongside off-chain computation, enabling hybrid smart contracts that combine on-chain logic with off-chain processing~\cite{chainlink2021v2}. Key innovations include Off-Chain Reporting (OCR) for reduced gas costs and super-linear staking security, where the bribe required to corrupt the network grows faster than the sum of individual deposits. Chainlink dominates the oracle market, facilitating over \$25 trillion in transaction value and securing nearly \$100 billion in DeFi total value locked. However, its architecture is optimized for structured, objective data feeds, asset prices, temperature readings, and sports scores with official APIs. It is less well suited to resolving subjective or ambiguous prediction market outcomes.

\subsubsection{Kalshi's Centralized Resolution}

Kalshi, the first CFTC-regulated Designated Contract Market for event contracts, represents the opposite end of the oracle spectrum from Chainlink's automated approach. An internal operations team resolves markets by verifying outcomes against pre-specified authoritative sources such as official government statistics, league results, and regulatory announcements, with an Outcome Review Committee for ambiguous cases~\cite{kalshi_outcomes}. This provides fast, reliable resolution with clear accountability. Kalshi's resolution decisions carry the weight of a regulated financial institution. However, the approach introduces single-entity trust assumptions and does not scale to the thousands of simultaneous markets that decentralized platforms support. Kalshi's resolution process is particularly relevant to this thesis because KalshiBench, our evaluation benchmark, is derived from Kalshi markets with verified outcomes. The ground truth labels therefore reflect Kalshi's authoritative resolution judgments~\cite{nel2025kalshibench}.

\subsubsection{UMA's Optimistic Oracle}

UMA Protocol operates on an assume correct unless challenged principle~\cite{uma2024oracle}. Asserters submit bonded assertions. If undisputed during a configurable liveness period of 2 hours to 2 days, assertions are accepted automatically. Only approximately 1.5\% of proposals are disputed. Disputes escalate to the Data Verification Mechanism (DVM), a commit-reveal vote by UMA token holders that resolves within 48 to 96 hours, with incorrect voters slashed. The system is designed so that corrupting the oracle by acquiring a supermajority of UMA tokens would cost more than any potential profit from corruption. Polymarket uses UMA's optimistic oracle for resolution of billions of dollars in trading volume~\cite{polymarket_uma_adapter}. The optimistic oracle's design embodies a natural tiered architecture consisting of cheap automation in the common case and expensive human arbitration in the rare disputed case. This structure directly informs our escalation framework.

\subsubsection{The Automation--Arbitration Tradeoff}

The fundamental tension in oracle design is between automation, which is fast, cheap, and scalable but limited to structured data and vulnerable to edge cases, and human arbitration, which handles ambiguity and subjectivity but is slow and expensive. As prediction markets grow in volume and complexity, particularly for subjective outcomes such as political events or regulatory decisions, neither pure automation nor human-only adjudication scales. This motivates the use of large language models as an intermediate resolution layer that can interpret natural-language resolution criteria, verify time-bounded events, and provide structured reasoning, while still deferring to human arbitration for genuinely ambiguous cases.

\subsection{LLMs as an Intermediate Resolution Layer}

The application of large language models to oracle resolution is a recent development driven by two capabilities. LLMs can interpret natural-language resolution criteria, such as "Will the Federal Reserve raise interest rates before March 2025?" Retrieval-augmented generation allows them to ground their reasoning in current evidence rather than relying solely on training data.

\subsubsection{The Forecasting--Resolution Distinction}

A critical conceptual distinction for this thesis is that forecasting and resolution impose different requirements on AI systems. Forecasting requires probabilistic reasoning about uncertain future events, where calibration, defined as the alignment between stated probabilities and empirical frequencies, is the central evaluation criterion. Oracle resolution, by contrast, requires determining whether a past event did occur, given available evidence and specific resolution criteria. Resolution is architecturally closer to fact verification, as in the FEVER benchmark~\cite{thorne2018fever}, which evaluates claim-level evidence retrieval and binary classification, than to probabilistic forecasting. Our thesis bridges both modes by using KalshiBench questions that have already resolved and evaluating models' ability to determine what happened given retrieved evidence rather than to predict what will happen.

\subsubsection{LLM Failure Modes Relevant to Oracle Resolution}

Several failure modes directly threaten oracle accuracy and motivate the 
move from single-LLM to multi-agent architectures. Hallucination, 
defined as generating plausible but incorrect outputs, has been extensively 
studied, with a key distinction between intrinsic hallucination, which 
contradicts the source, and extrinsic hallucination, which is unverifiable 
from the source~\cite{ji2023hallucination}. RLHF training can exacerbate 
hallucination by rewarding plausible-sounding but incorrect answers, and 
RAG alone is insufficient to prevent it~\cite{huang2025hallucination}. 
Sycophancy, the tendency to agree with social pressure rather than 
reason independently, is particularly concerning for multi-agent designs: 
sycophantic behavior has been observed in 58\% of cases across multiple 
frontier models, with a 78.5\% persistence rate once 
triggered~\cite{fanous2025sycophancy}. In multi-agent systems, sycophancy 
could cause agents to agree with majority positions rather than reasoning 
independently, directly undermining the diversity assumption underlying 
aggregation benefits. Temporal reasoning errors, identified as the 
primary bottleneck by both Chainlink and UMA systems, represent a 
category-specific challenge where LLMs struggle with date arithmetic, 
relative time expressions, and determining whether evidence predates or 
postdates an event of interest.

These failure modes, including hallucination without self-correction, sycophantic convergence, and temporal confusion, cannot be addressed by single-model improvements alone. They motivate the multi-agent architectures evaluated in this thesis, where cross-model critique and independent verification can potentially catch errors that a single model would propagate unchecked.

\section{Methodology}
\label{methodology}
\begin{figure}[htbp]
    \centering
    \includegraphics[width=0.9\linewidth]{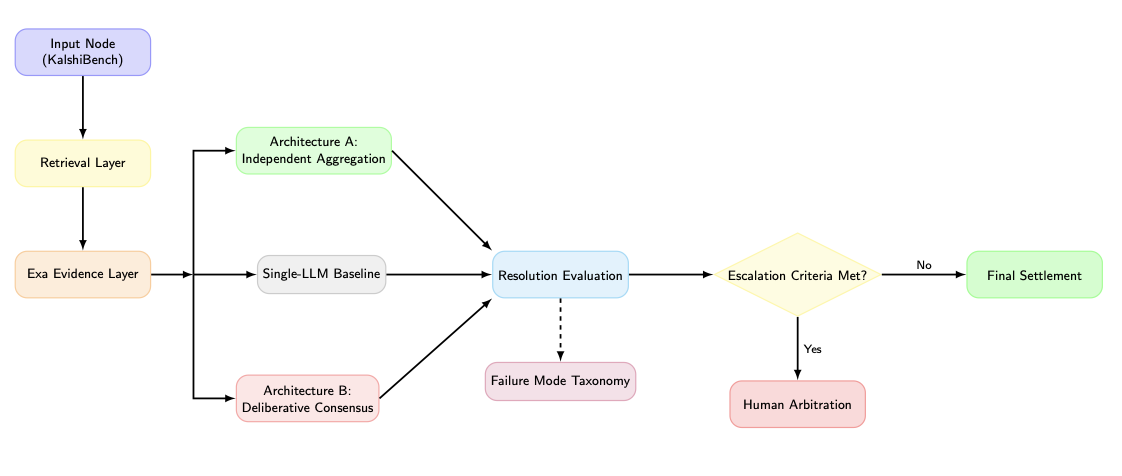}
    \caption{System architecture for multi-agent oracle resolution.}
    \label{fig:architecture}
\end{figure}

This section describes the system architecture, dataset, retrieval pipeline, multi-agent resolution architectures, and evaluation framework used in this study. The overall pipeline is shown in Figure~\ref{fig:architecture}.

\subsection{Evaluation Dataset}

We evaluate on KalshiBench v2, a publicly available benchmark of prediction market questions from Kalshi with verified real-world outcomes \cite{nel2025kalshibench}. Each record contains a question, natural-language resolution criteria, a category label, a close time, and a binary ground-truth outcome (YES or NO).

\begin{table}[t]
\centering
\caption{Summary statistics for the filtered KalshiBench v2 evaluation dataset (markets closed after May 2025).}
\label{tab:dataset-stats}
\begin{tabular}{lr}
\toprule
\textbf{Statistic} & \textbf{Value} \\
\midrule
Total questions & 1,189 \\
YES outcomes & 488 (41.0\%) \\
NO outcomes & 701 (59.0\%) \\
Number of categories & 16 \\
Filter criterion & Markets closed after May 2025 \\
\bottomrule
\end{tabular}
\vspace{0.5em}
\end{table}

\begin{table}[t]
\centering
\caption{Category distribution in the filtered KalshiBench v2 evaluation dataset (1,189 questions).}
\label{tab:category-distribution}
\begin{tabular}{lrr}
\toprule
\textbf{Category} & \textbf{Count} & \textbf{Percentage} \\
\midrule
Politics & 311 & 26.16\% \\
Sports & 269 & 22.62\% \\
Entertainment & 230 & 19.34\% \\
Elections & 80 & 6.73\% \\
Crypto & 59 & 4.96\% \\
Mentions & 52 & 4.37\% \\
Companies & 50 & 4.21\% \\
Financials & 30 & 2.52\% \\
World & 29 & 2.44\% \\
Climate and Weather & 27 & 2.27\% \\
Economics & 21 & 1.77\% \\
Social & 13 & 1.09\% \\
Science and Technology & 10 & 0.84\% \\
Transportation & 4 & 0.34\% \\
Education & 2 & 0.17\% \\
Health & 2 & 0.17\% \\
\bottomrule
\end{tabular}
\end{table}

To reduce the risk of training-data contamination, we filter the dataset to include only markets that closed on or after May 1, 2025, a date chosen to fall beyond the training cutoff of the frontier models used in our experiments. After filtering, the evaluation set contains 1,189 questions, with 488 YES outcomes (41.0\%) and 701 NO outcomes (59.0\%). The dataset spans 16 categories, with Politics, Sports, and Entertainment accounting for the majority of questions and the remaining 13 categories providing broad topical coverage for category-level analysis.

\subsection{Evidence Retrieval Layer}

All agents receive identical evidence for each question. This shared-evidence control isolates differences in reasoning capability from differences in information access.

We use Exa as the retrieval backend. For each question, the system constructs a search query using the template \texttt{"What is the result of this question \allowbreak <question\_text>"} and retrieves up to 10 sources. A temporal filter restricts results to documents published on or before one day after the market's resolution date, ensuring the evidence set reflects information that would have been available at the time of resolution.

Retrieved content is packaged into a structured \texttt{EvidencePacket} containing the question metadata, retrieval timestamp, query text, and a list of sources with titles, URLs, publication dates, and extracted highlights. The same packet is passed verbatim to every agent in both architectures. Evidence packets are cached locally by retrieval mode and question ID to ensure deterministic reruns and reduce API cost across experimental iterations.

\subsection{LLM Agent Design}

We employ three LLM agents selected for architectural diversity: GPT-5 nano (OpenAI), DeepSeek-V3 (DeepSeek, served via Together AI), and Llama-3.3-70B-Instruct-Turbo (Meta, served via Together AI). This selection spans a proprietary frontier model, a large open-weights reasoning model, and an open-weights instruction-tuned model, providing diversity in training data, architecture, and optimization approach.

Each agent is implemented as a subclass of a shared base class that enforces a standardized output schema. For every question, each agent returns a binary decision (YES or NO), a confidence score between 0 and 1, a natural-language reasoning trace, latency in milliseconds, and token usage statistics. If an agent encounters an error, it returns a structured failure record rather than crashing the pipeline, allowing partial results to be analyzed.

All agents receive the same system and user prompt templates. The system prompt establishes the agent's role as a prediction market oracle resolver, and the user prompt injects the question text, resolution criteria, resolution date, and the formatted evidence packet. Structured output is enforced through provider-specific mechanisms. Prompt templates for the single-agent baseline (used in Architecture A and the standalone baselines) are provided in Appendix Figure~\ref{fig:system-prompt}. The deliberation prompts used in Architecture B are provided in Appendix Figures~\ref{fig:archB-round1-prompt} and~\ref{fig:archB-round2-prompt}.

The OpenAI agent uses JSON schema enforcement, while the Together-hosted models (DeepSeek and Llama) attempt JSON schema enforcement and fall back to JSON object mode if the schema is rejected. All agents use a retry strategy of up to 3 attempts with exponential backoff.

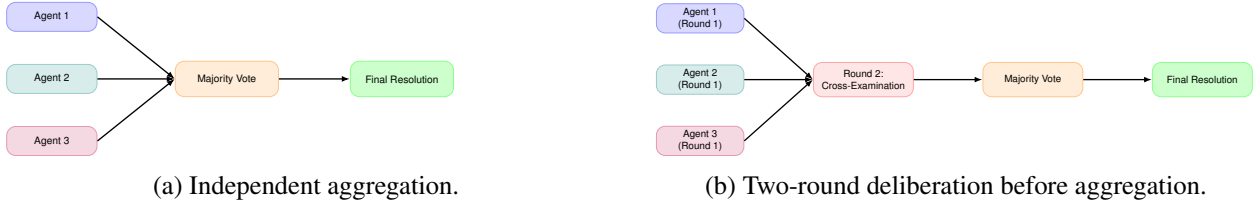
\begin{figure*}[t]
\centering

\begin{subfigure}[t]{0.48\textwidth}
\centering
\resizebox{\linewidth}{!}{%
\begin{tikzpicture}[
    node distance=1.2cm and 2.0cm,
    box/.style={
        draw, thick,
        rounded corners=8pt,
        minimum width=3.2cm, minimum height=1.1cm,
        align=center, font=\footnotesize\sffamily
    },
    smallbox/.style={
        draw, thick,
        rounded corners=8pt,
        minimum width=2.8cm, minimum height=0.9cm,
        align=center, font=\footnotesize\sffamily
    },
    arrow/.style={-{Latex[length=2.5mm]}, very thick}
]

\node[smallbox, fill=blue!15, draw=blue!40] (a1) at (-3.8,1.2) {Agent 1};
\node[smallbox, fill=teal!15, draw=teal!40, below=1.0cm of a1] (a2) {Agent 2};
\node[smallbox, fill=purple!15, draw=purple!40, below=1.0cm of a2] (a3) {Agent 3};

\node[box, fill=orange!15, draw=orange!45, right=2.4cm of a2] (vote) {Majority Vote};
\node[box, fill=green!20, draw=green!45, right=2.2cm of vote] (final) {Final Resolution};

\draw[arrow] (a1.east) -- (vote.west);
\draw[arrow] (a2.east) -- (vote.west);
\draw[arrow] (a3.east) -- (vote.west);
\draw[arrow] (vote.east) -- (final.west);

\node[inner sep=0pt, outer sep=0pt] (padA) at ($(final.east)+(4.6,0)$) {};

\end{tikzpicture}
}%
\caption{Independent aggregation.}
\label{fig:archA-diagram}
\end{subfigure}
\hfill
\begin{subfigure}[t]{0.48\textwidth}
\centering
\resizebox{\linewidth}{!}{%
\begin{tikzpicture}[
    node distance=1.2cm and 2.0cm,
    box/.style={
        draw, thick,
        rounded corners=8pt,
        minimum width=3.2cm, minimum height=1.1cm,
        align=center, font=\footnotesize\sffamily
    },
    smallbox/.style={
        draw, thick,
        rounded corners=8pt,
        minimum width=2.8cm, minimum height=0.9cm,
        align=center, font=\footnotesize\sffamily
    },
    arrow/.style={-{Latex[length=2.5mm]}, very thick}
]

\node[smallbox, fill=blue!15, draw=blue!40] (b1) at (-4.0,1.2) {Agent 1\\(Round 1)};
\node[smallbox, fill=teal!15, draw=teal!40, below=1.0cm of b1] (b2) {Agent 2\\(Round 1)};
\node[smallbox, fill=purple!15, draw=purple!40, below=1.0cm of b2] (b3) {Agent 3\\(Round 1)};

\node[box, fill=red!10, draw=red!35, right=2.2cm of b2] (cross) {Round 2:\\Cross-Examination};
\node[box, fill=orange!15, draw=orange!45, right=2.2cm of cross] (vote) {Majority Vote};
\node[box, fill=green!20, draw=green!45, right=2.2cm of vote] (final) {Final Resolution};

\draw[arrow] (b1.east) -- (cross.west);
\draw[arrow] (b2.east) -- (cross.west);
\draw[arrow] (b3.east) -- (cross.west);
\draw[arrow] (cross.east) -- (vote.west);
\draw[arrow] (vote.east) -- (final.west);

\end{tikzpicture}
}%
\caption{Two-round deliberation before aggregation.}
\label{fig:archB-diagram}
\end{subfigure}

\caption{Resolution architectures evaluated in this study. Architecture A aggregates independent agent decisions. Architecture B adds a cross-examination round before aggregation.}
\label{fig:archA-archB-panels}
\end{figure*}

\subsection{Architecture A: Independent Aggregation}

In Architecture A, each of the three agents resolves every question independently and in parallel. No agent has access to the decisions or reasoning of the others. The system then aggregates their binary decisions by majority vote. The final outcome is YES if at least two of the three agents vote YES, and NO otherwise. If one or more agents fail (returning no valid decision), the vote is computed over the remaining successful responses. If the vote is tied, the system defaults to NO.

This architecture mirrors the logic of decentralized oracle networks such as Chainlink, which aggregate independent data feeds from multiple node operators.

\subsection{Architecture B: Deliberative Consensus}

Architecture B extends the independent setup with a structured deliberation phase. Resolution proceeds in two rounds. A structural comparison of Architectures A and B is shown in Figure~\ref{fig:archB-diagram}.

In Round 1, each agent independently resolves the question, producing a decision, confidence score, and reasoning trace, identical to the process in Architecture A.

In Round 2, each agent receives the anonymized reasoning and decisions from the other two agents alongside the original evidence. Agents are prompted to consider the peer arguments and may revise their position or reaffirm their original decision. Each agent produces a new decision, confidence score, and updated reasoning trace.

The final aggregation uses a tiered rule. The system first checks for a majority among the Round 2 decisions. If Round 2 produces a tie, the system falls back to the Round 1 majority. If both rounds tie, the system defaults to NO. The tie-break method used is recorded in the output for every question. A tie can only occur when an API failure reduces the agent pool to two.

\subsection{Single-LLM Baselines}

To contextualize the multi-agent results, we evaluate each of the three models as a standalone resolver. Each baseline uses the same evidence retrieval, prompting, and structured output pipeline as the multi-agent architectures. The only difference is that a single model's decision is taken as the final answer with no aggregation. This allows us to measure the marginal accuracy gain (or loss) from multi-agent aggregation and deliberation relative to the best, worst, and average individual model.

\subsection{Evaluation Metrics}

We evaluate the system along several dimensions. First, we measure overall resolution accuracy against the ground-truth outcome for each question, reporting both aggregate performance and category-level breakdowns across single-LLM baselines, Architecture A, and Architecture B. Second, we analyze inter-agent agreement rates to characterize how often the three models independently converge on the same answer and how frequently aggregation changes the final outcome. Third, for the deliberative architecture, we measure convergence dynamics across rounds, including how often agents revise their decisions, whether revisions improve or degrade correctness, and the rate at which deliberation produces unanimous consensus.

\subsection{Failure Mode Taxonomy}

To characterize the qualitative nature of resolution errors, we define five failure types based on inspection of incorrect resolutions: retrieval failure, temporal reasoning error, hallucination, shared bias, and persuasive error propagation. We use the taxonomy to identify and illustrate distinct mechanisms by which the system fails, with representative examples drawn from the reasoning traces and evidence packets of incorrect resolutions.

\subsection{Escalation Analysis}

To inform the design of hybrid AI-human oracle systems, we analyze whether signals observable at resolution time can reliably distinguish questions the system will answer correctly from those it will get wrong. We define two escalation signals derived from the multi-agent output of both architectures. The first is inter-agent agreement: whether all three models produced the same binary decision (unanimous) or whether the vote was 2-1 (split). The second is average confidence: the arithmetic mean of the three models' self-reported confidence scores for each question. We combine these signals into a composite escalation score defined as:

\[
\text{composite\_score} = \mathbf{1}[\text{unanimous}] + \overline{\text{confidence}}
\]

This score ranges from 0 to 2, with higher values indicating greater system certainty. To evaluate the practical utility of these signals, we construct a coverage-accuracy tradeoff curve by ranking all questions in descending order of composite score and computing cumulative accuracy at each coverage level, where coverage is the fraction of questions auto-resolved rather than escalated to human arbitration. This analysis follows the selective prediction framework introduced by \cite{chow1970optimum} and later formalized by \cite{geifman2017selective}, in which a system that declines to answer uncertain questions achieves higher accuracy on those it does answer, at the cost of resolving fewer questions automatically.

\subsection{Implementation}

The system is implemented in Python, with modules for data loading, 
evidence retrieval, agent orchestration, and result aggregation. All 
per-question outputs and cached evidence packets are retained to 
support reproducibility; command-line entry points and exact 
invocations are documented in Appendix~\ref{sec:cli-reference}.

\section{Results}
We evaluated two multi-agent oracle architectures on 1,189 resolved prediction market questions from KalshiBench. Three LLMs served as oracle agents: GPT-5 Nano, DeepSeek-V3, and 
Llama-3.3-70B. All agents received identical retrieved evidence for each question, isolating reasoning capability from retrieval quality. Architecture A (Independent Aggregation) has each model resolve independently, with outcomes determined by majority vote or confidence-weighted vote. Architecture B (Deliberative Consensus) has models engage in structured debate over two rounds, viewing each other's reasoning before submitting final answers.

\subsection{Overall Accuracy: Independent Aggregation Outperforms Both Single-Model Baselines and Deliberative Consensus}

\begin{table}[t]
\centering
\begin{tabular}{llc}
\toprule
\textbf{Group} & \textbf{System} & \textbf{Accuracy (\%)} \\
\midrule
\multirow{3}{*}{Single-LLM Baselines}
    & GPT-4o          & 82.34 \\
    & DeepSeek        & 82.42 \\
    & Llama           & 81.24 \\
\midrule
\multirow{2}{*}{Architecture A}
    & Majority Vote            & 82.93 \\
    & \textbf{Confidence-Weighted Vote} & \textbf{83.43} \\
\midrule
\multirow{4}{*}{Architecture B}
    & Final-Round GPT-4o       & 74.35 \\
    & Final-Round DeepSeek     & 76.70 \\
    & Final-Round Llama        & 76.37 \\
    & Deliberative Final       & 76.11 \\
\midrule
Prior Work (KalshiBench)
    & Claude Opus 4.5          & 69.3 \\
\bottomrule
\end{tabular}
\caption{Overall resolution accuracy across all 1,189 questions.}
\label{tab:overall}
\end{table}

\begin{figure}[t]
\centering
\includegraphics[width=\linewidth]{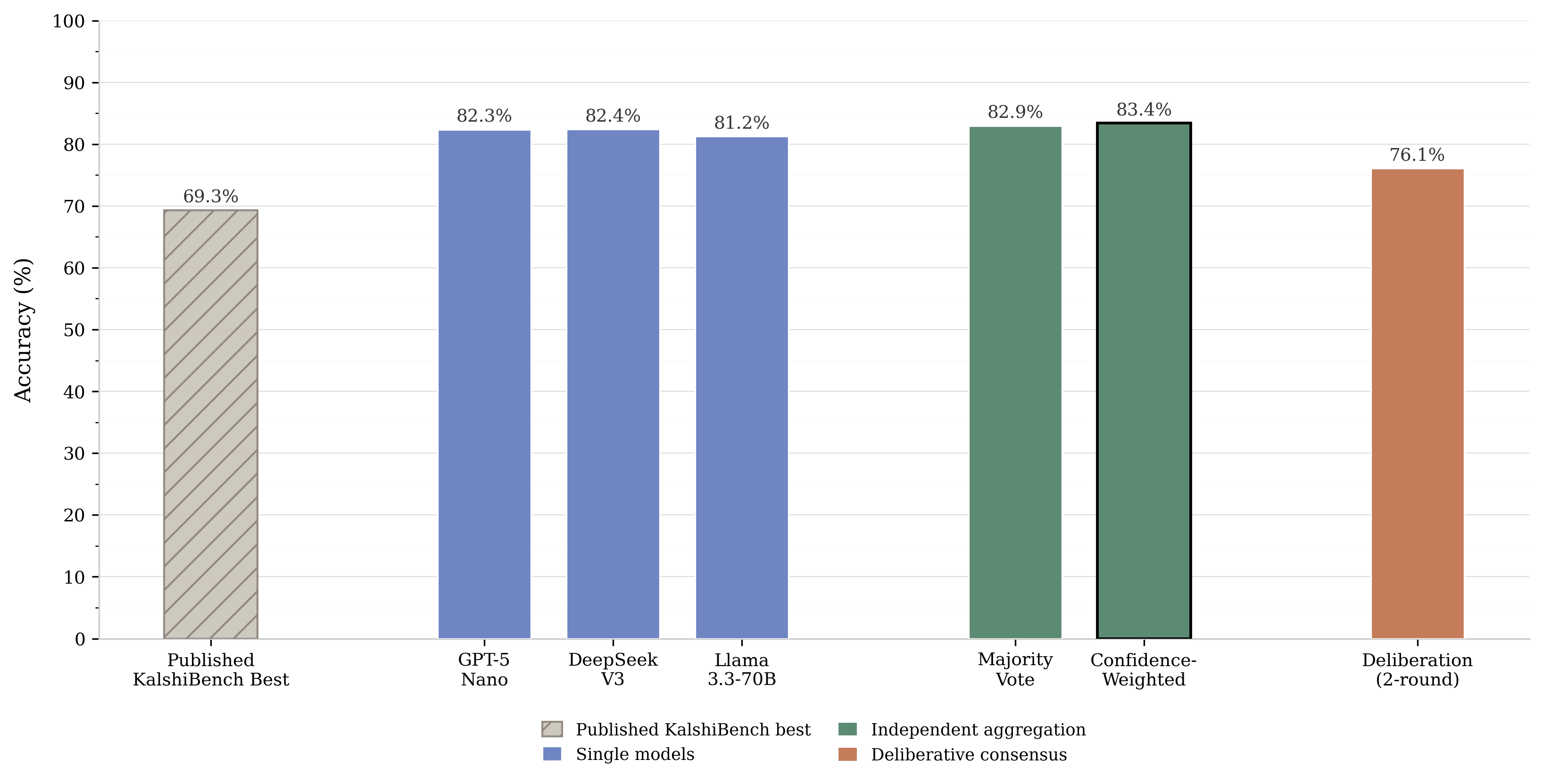}
\caption{\textbf{Overall resolution accuracy on 1,189 KalshiBench questions.} Confidence-weighted independent aggregation (Architecture A) achieves the highest accuracy at 83.43\%, outperforming the best single-model baseline by 1.01 percentage points. Deliberative consensus (Architecture B) degrades accuracy to 76.11\%, below all single-model baselines. The hatched bar shows the best previously published KalshiBench result}
\label{fig:overall_accuracy}
\end{figure}

Table~\ref{tab:overall} presents overall resolution accuracy across all 1,189 questions. Architecture A's confidence-weighted vote achieves the highest accuracy at 83.43\%, representing a +1.01 percentage point improvement over the best individual model (DeepSeek, 82.42\%). Majority voting also outperforms every single-model baseline at 82.93\%. Architecture B's deliberative process, by contrast, \emph{degrades} performance substantially: all final-round outputs fall below every single-model baseline, with the best deliberative result (Final-Round DeepSeek, 76.70\%) trailing the worst individual model (Llama, 81.24\%) by over four percentage points. As shown in Figure~\ref{fig:overall_accuracy}, independent aggregation consistently outperforms both single-model baselines and deliberative consensus.

All systems substantially outperform the best previously reported result on KalshiBench, Claude Opus 4.5 at 69.3\%, though this comparison should be interpreted cautiously, as our pipeline uses a different retrieval layer (Exa with date-filtered evidence) than the original KalshiBench evaluation (which relied on the Model's general web search API). The improvement likely reflects the combined effect of better retrieval and multi-agent reasoning rather than model superiority alone.

To confirm that Architecture A's advantage over Architecture B is not an artifact of variance, we conducted McNemar's test for paired differences and a head-to-head comparison on the 1,170 overlapping questions evaluated by both architectures (Table~\ref{tab:mcnemar} and Table~\ref{tab:h2h}).

\begin{table}[t]
\centering
\caption{McNemar's test for paired accuracy differences. Each row compares two systems on the same question set. ``A only'' and ``B only'' indicate the number of questions each system answered correctly that the other missed. Significant results ($p < 0.05$) are bolded.}
\label{tab:mcnemar}
\begin{tabular}{llcccc}
\toprule
\textbf{System A} & \textbf{System B} & \textbf{A only} & \textbf{B only} & \textbf{Discordant} & $p$\textbf{-value} \\
\midrule
A Majority & A Weighted & 1 & 7 & 8 & 0.070 \\
\textbf{A Majority} & \textbf{B Deliberative} & \textbf{117} & \textbf{37} & \textbf{154} & $\mathbf{7.2 \times 10^{-11}}$ \\
\textbf{A Weighted} & \textbf{B Deliberative} & \textbf{123} & \textbf{37} & \textbf{160} & $\mathbf{5.5 \times 10^{-12}}$ \\
\bottomrule
\end{tabular}
\end{table}

Architecture A significantly outperforms Architecture B ($p < 10^{-10}$), and the asymmetry is stark: Architecture A correctly resolves 117 questions that B gets wrong, while B captures only 37 that A misses, a roughly 3:1 ratio. Within Architecture A, the difference between majority and confidence-weighted voting is not statistically significant ($p = 0.07$), though weighted voting trends better by picking up 7 additional correct resolutions at the cost of only 1.

\begin{table}[t]
\centering
\caption{Head-to-head comparison between Architecture A (confidence-weighted vote) and Architecture B (deliberative final) on the $n = 1{,}170$ overlapping questions evaluated by both systems. ``A correct, B wrong'' indicates questions where only independent aggregation produced the correct resolution.}
\label{tab:h2h}
\begin{tabular}{lc}
\toprule
\textbf{Outcome} & \textbf{Count (\%)} \\
\midrule
Both correct       & 852 (72.8\%) \\
A correct, B wrong & 117 (10.0\%) \\
B correct, A wrong &  37 (3.2\%) \\
Both wrong         & 164 (14.0\%) \\
\midrule
Architecture A accuracy & 82.8\% \\
Architecture B accuracy & 76.0\% \\
\bottomrule
\end{tabular}
\end{table}

The 164 questions (14.0\%) wrong under both architectures represent a ``hard core'' of failures that multi-agent reasoning alone cannot resolve. These questions likely require either higher-quality evidence retrieval or human judgment, a finding we revisit in the Escalation Analysis (Section~\ref{sec:escalation}).

\subsubsection{Robustness Check: Unique-Market Accuracy and Multi-Instance Consistency}

Because KalshiBench contains multi-instance markets, where a single market (e.g., ``Which company will IPO in 2026?'') is decomposed into multiple binary rows, one per possible outcome (e.g., ``Did Stripe IPO in 2026?'', ``Did Klarna IPO in 2026?''), the 1,189 rows correspond to only 703 unique markets. Of these, 486 markets appear in more than one row. To ensure our headline results are not inflated by over-counting repeated instances, we also evaluated accuracy at the unique-market level, keeping one row per \texttt{question\_id}. Architecture A achieves 81.9\% accuracy on unique markets (95\% Wilson CI: [0.789, 0.846]), and Architecture B achieves 75.5\% (95\% Wilson CI: [0.722, 0.786]). These numbers are consistent with the full-dataset results, confirming that the accuracy gap between architectures is not an artifact of repeated market weighting.

The multi-instance markets also reveal an important finding about resolution \emph{consistency}. For these 486 markets, each architecture resolves multiple binary outcomes drawn from the same parent market with the same retrieved evidence. Table~\ref{tab:consistency} shows how often each architecture produces consistent outcomes across all instances of a given market.

\begin{table}[t]
\centering
\caption{Resolution consistency across 486 multi-instance markets (markets with more than one binary outcome row in the dataset). ``All correct'' means every outcome for that market was resolved correctly. ``Mixed'' means at least one outcome was correct and at least one was incorrect within the same market, despite shared underlying evidence. ``All incorrect'' means every outcome was resolved incorrectly.}
\label{tab:consistency}
\begin{tabular}{lccc}
\toprule
\textbf{Architecture} & \textbf{All Correct} & \textbf{Mixed} & \textbf{All Incorrect} \\
\midrule
A (Independent)  & 380 (78.2\%) & 67 (13.8\%) & 39 (8.0\%) \\
B (Deliberative) & 338 (69.5\%) & 90 (18.5\%) & 58 (11.9\%) \\
\bottomrule
\end{tabular}
\end{table}

Architecture A resolves all outcomes correctly for 78.2\% of multi-instance markets compared to 69.5\% for Architecture B. More notably, Architecture B produces mixed-consistency outcomes on 90 markets (18.5\%) versus 67 (13.8\%) for Architecture A. These mixed cases are particularly concerning for oracle design: given the same parent market and the same retrieved evidence, the system correctly resolves some outcomes but not others. In a deployed oracle, this inconsistency could create exploitable discrepancies across related contracts within the same market. The finding suggests that deliberation not only reduces accuracy but also introduces additional variance into the resolution process, likely because the multi-round debate amplifies sensitivity to how individual outcomes within a market are framed.

These results partially align with prior multi-agent literature. The Iterative Consensus Ensemble (ICE) framework reported 7--15 percentage point improvements from multi-agent aggregation on medical reasoning benchmarks~\cite{omar2025ice}. Our independent aggregation gains are more modest (+1.01 pp), likely because our single-model baselines already achieve over 82\% accuracy, leaving less room for improvement. More critically, our deliberation architecture reduces accuracy, the opposite of what Du et al.~\cite{du2023debate} found for arithmetic tasks, where three-agent debate improved accuracy from roughly 70\% to 95\%. The difference likely reflects the nature of oracle resolution: unlike arithmetic where a correct derivation can be verified step-by-step, prediction market outcomes depend on evidence interpretation where confident-sounding but incorrect reasoning can be persuasive enough to flip correct initial judgments, a dynamic we analyze in Section~\ref{sec:errcorr}.

\subsection{Category-Level Performance Reveals Consistent Gains on Well-Covered 
Domains and Fundamental Difficulty in Ambiguous Ones}
\label{sec:category}

\begin{figure}[t]
\centering
\includegraphics[width=0.65\textwidth]{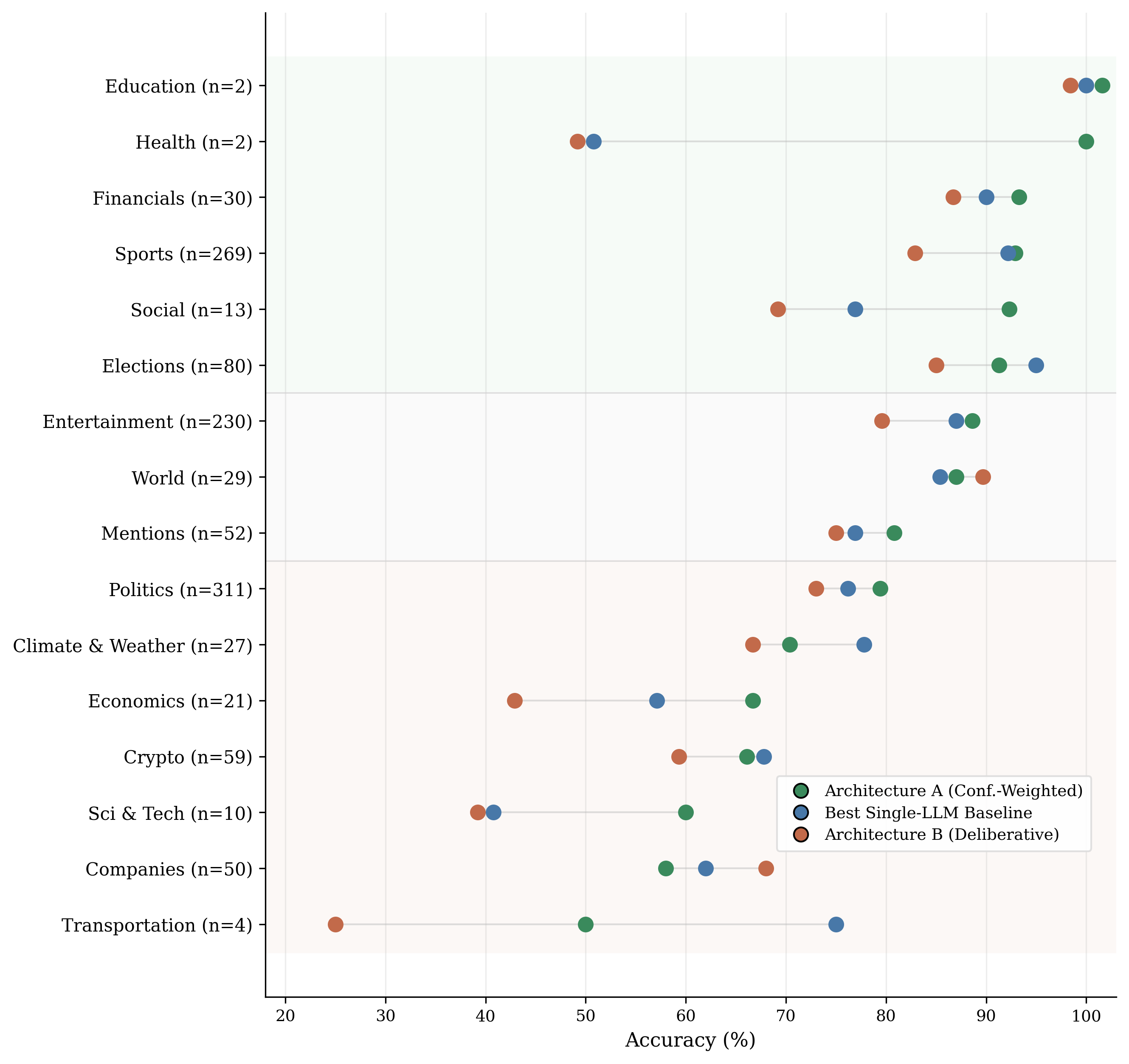}
\caption{\textbf{Category-level accuracy for Architecture A (confidence-weighted), best single-LLM baseline, and Architecture B (deliberative).} Categories are sorted by Architecture A accuracy. Alternating shaded bands delineate the tier structure discussed in the text. Categories with $n < 10$ should be interpreted with caution.}
\label{fig:category_dotplot}
\end{figure}

Table~\ref{tab:category} breaks down accuracy by question category. Categories are sorted by Architecture A's confidence-weighted accuracy, with disagreement rate (the percentage of questions where at least one model dissented from the majority) shown in the rightmost column. Figure~\ref{fig:category_dotplot} visualizes this breakdown.

\begin{table}[t]
\centering
\small
\begin{tabular}{lrccccc}
\toprule
\textbf{Category} & $n$ & \textbf{A Wtd} & \textbf{A Maj} & \textbf{Best Baseline} & \textbf{B Delib} & \textbf{Disagree \%} \\
\midrule
Education           &   2 & 100.0 & 100.0 & 100.0 & 100.0 &  0.0 \\
Health              &   2 & 100.0 & 100.0 &  50.0 &  50.0 & 50.0 \\
Financials          &  30 &  93.3 &  93.3 &  90.0 &  86.7 & 13.3 \\
Sports              & 269 &  92.9 &  91.4 &  92.2 &  82.9 & 13.0 \\
Social              &  13 &  92.3 &  92.3 &  76.9 &  69.2 & 30.8 \\
Elections           &  80 &  91.3 &  91.3 &  95.0 &  85.0 & 13.8 \\
Entertainment       & 230 &  87.8 &  87.8 &  87.8 &  79.6 & 11.7 \\
World               &  29 &  86.2 &  86.2 &  86.2 &  89.7 &  6.9 \\
Mentions            &  52 &  80.8 &  80.8 &  76.9 &  75.0 & 19.2 \\
Politics            & 311 &  79.4 &  79.1 &  76.2 &  73.0 & 24.4 \\
Climate \& Weather  &  27 &  70.4 &  74.1 &  77.8 &  66.7 & 25.9 \\
Economics           &  21 &  66.7 &  66.7 &  57.1 &  42.9 & 28.6 \\
Crypto              &  59 &  66.1 &  62.7 &  67.8 &  59.3 & 18.6 \\
Sci \& Tech         &  10 &  60.0 &  60.0 &  40.0 &  40.0 & 40.0 \\
Companies           &  50 &  58.0 &  58.0 &  62.0 &  68.0 & 20.0 \\
Transportation      &   4 &  50.0 &  50.0 &  75.0 &  25.0 & 75.0 \\
\bottomrule
\end{tabular}
\caption{Accuracy by category for the top-performing systems. $n$ = number of questions per category. A~Wtd = Architecture A confidence-weighted vote. A~Maj = Architecture A majority vote. Best Baseline = highest single-model accuracy in each category. B~Delib = Architecture B deliberative final. Disagree~\% = proportion of questions where at least one of the three models produced a different answer from the other two.}
\label{tab:category}
\end{table}

A clear tier structure emerges across categories. The top tier includes Sports ($n=269$, 92.9\%), Financials ($n=30$, 93.3\%), and Elections ($n=80$, 91.3\%). These are well-covered domains where retrieved evidence tends to be unambiguous and authoritative. In these categories, inter-model disagreement is low (11--14\%), and Architecture A either matches or slightly exceeds the best single-model baseline. This pattern is consistent with Chainlink's findings, which reported 99.7\% accuracy on sports questions where outcomes are well-documented in public sources~\cite{chainlink2025aioracles}. Our somewhat lower sports accuracy (92.9\% vs.\ 99.7\%) may reflect that KalshiBench includes more temporally nuanced sports questions, such as season-level outcomes and conditional events, compared to Chainlink's simpler game-result format.

The middle tier includes Politics ($n=311$, 79.4\%), Mentions ($n=52$, 80.8\%), and Entertainment ($n=230$, 87.8\%). These categories show moderate difficulty with higher disagreement rates (12--24\%). In Politics, which constitutes the largest single category, Architecture A's weighted vote outperforms the best single baseline by 3.2 percentage points (79.4\% vs.\ 76.2\%), suggesting that aggregation provides meaningful benefit precisely when questions involve interpretive ambiguity. This aligns with the theoretical motivation for multi-agent approaches: diverse models are most valuable when no single model reliably dominates.

The bottom tier includes Crypto ($n=59$, 66.1\%), Companies ($n=50$, 58.0\%), and Economics ($n=21$, 66.7\%). These categories exhibit low accuracy across \emph{all} systems, indicating that these failures stem from fundamental difficulty rather than any architecture-specific weakness. Crypto and Companies questions often involve rapidly evolving market conditions, ambiguous resolution criteria (e.g., ``Will token X reach price Y by date Z?''), and thin or contradictory evidence. These are challenges that neither additional models nor deliberation can overcome if the underlying evidence is insufficient. Notably, Companies is the one category where Architecture B (68.0\%) meaningfully outperforms Architecture A (58.0\%), suggesting that deliberation may occasionally help on information-rich questions where models benefit from synthesizing multiple interpretive frames. World events show a similar but smaller pattern (B at 89.7\% vs.\ A at 86.2\%).

One finding that merits caution is Elections. It shows an unusual case where the best single-model baseline (95.0\%) exceeds Architecture A (91.3\%). Manual inspection revealed that one model occasionally held the correct minority position on election questions, and aggregation overrode it. This illustrates a known limitation of majority-rule systems: when one model is systematically better-calibrated on a specific domain, aggregation can dilute its advantage. A potential mitigation would be category-aware weighting, where models receive different influence based on domain-specific track records, though this introduces the risk of overfitting to historical category performance.

Several categories (Education $n=2$, Health $n=2$, Transportation $n=4$, Sci \& Tech $n=10$) have sample sizes too small to draw reliable conclusions. We report their results for completeness but caution against over-interpreting apparent patterns in these categories.

\subsection{Correlated Errors: Why Aggregation Gains Fall Short of the Theoretical Ceiling}
\label{sec:errcorr}

The central promise of multi-agent architectures is that diverse models will make \emph{different} errors, allowing aggregation to filter out individual mistakes. To test this premise, we computed pairwise error correlations and conditional error probabilities across the three oracle agents (Tables~\ref{tab:errcorr} and~\ref{tab:pairwise}).

Error correlations range from moderate to high ($r = 0.529$--$0.689$). GPT-4o exhibits the lowest correlation with the other two models ($r = 0.53$ with Llama, $r = 0.58$ with DeepSeek), making it the most independently-minded oracle agent. DeepSeek and Llama are the most correlated pair ($r = 0.689$), consistent with the hypothesis that open-source models sharing overlapping training data or architectural lineage may inherit similar reasoning blind spots. The conditional error probabilities reinforce this pattern: when DeepSeek is wrong, Llama is also wrong 77.0\% of the time, compared to 63.3\% for GPT-4o when Llama errs.

These correlations have direct implications for the theoretical ceiling of aggregation-based approaches. Under an independence assumption, three models each achieving $\sim$82\% accuracy would be expected to reach approximately 91\% accuracy under majority voting according to the Condorcet Jury Theorem. However, the observed majority-vote accuracy of 82.93\% falls substantially short of this theoretical ceiling. This gap is consistent with the moderate-to-high error correlations observed across models: when one model fails, the others are often failing on the same question for the same underlying reason, typically because the retrieved evidence is ambiguous, incomplete, or misleading.

\begin{table}[t]
\centering
\caption{Pearson correlation of binary error vectors across oracle agents. Higher values indicate models tend to make mistakes on the same questions. A correlation of 1.0 would mean perfectly overlapping errors; 0.0 would mean fully independent error patterns.}
\label{tab:errcorr}
\begin{tabular}{lccc}
\toprule
 & \textbf{GPT-4o} & \textbf{DeepSeek} & \textbf{Llama} \\
\midrule
GPT-4o   & 1.000 & 0.580 & 0.529 \\
DeepSeek & 0.580 & 1.000 & 0.689 \\
Llama    & 0.529 & 0.689 & 1.000 \\
\bottomrule
\end{tabular}
\end{table}

These correlated errors manifest in two ways within our evaluation. First, they explain why Architecture~A's accuracy gains over single-model baselines are modest (+1.01 pp rather than the $\sim$9 pp that independence would predict). Second, they create the conditions under which Architecture~B's deliberation actively degrades performance, a dynamic we analyze as \emph{persuasive error propagation} in Section~\ref{sec:failmode-pep}.

\subsection{Failure Mode Taxonomy}
\label{sec:failure}

The preceding sections quantify accuracy differences across architectures and measure error correlations across models. This section complements those results by analyzing incorrect resolutions to understand why the system fails. We identify five failure modes, four shared across architectures and one specific to Architecture~B's deliberation process.

These failure modes differ primarily in where the error originates. Retrieval failures stem from missing evidence upstream of the resolution pipeline. Temporal reasoning errors arise from misinterpreting the time-relevance of available evidence. Hallucinations involve fabricated claims or criteria interpretations not grounded in the evidence packet. Shared bias reflects correlated misinterpretation of evidence that is present but systematically misread across all models. Finally, persuasive error propagation is unique to Architecture~B, emerging not from individual reasoning but from inter-agent interaction during deliberation.
\begin{table}[t]
\centering
\caption{Conditional error probability: $P(\text{B wrong} \mid \text{A wrong})$. Each cell shows the probability that model B also answers incorrectly, given that model A answered incorrectly. Values above 0.50 indicate positively correlated failure patterns.}
\label{tab:pairwise}
\begin{tabular}{lcc}
\toprule
\textbf{Given A wrong} & \textbf{B = DeepSeek} & \textbf{B = Llama} \\
\midrule
GPT-4o   & 0.652 & 0.633 \\
DeepSeek & ---   & 0.770 \\
Llama    & 0.722 & --- \\
\bottomrule
\end{tabular}
\end{table}

\subsubsection{Retrieval Failure}

A retrieval failure occurs when the evidence packet lacks the information needed to resolve the question correctly, so that agents reach the wrong conclusion despite sound reasoning. The error is upstream of the resolution pipeline, and no amount of multi-agent aggregation or deliberation can compensate.

A representative example is the question (Row 0189) ``If Glen Powell hosts an episode of Saturday Night Live Season 51, then the market resolves to Yes'' (ground truth: YES). The evidence packet contained ten sources listing confirmed SNL Season 51 hosts, but none mentioned Glen Powell. All three agents independently concluded NO, citing his absence from multiple credible host lists. The Exa query returned articles that were either published before Powell's hosting was announced or simply did not mention it, leaving the agents with a systematically incomplete evidence set.

Retrieval failures are particularly difficult to detect at resolution time because the evidence packet appears substantive, agents receive multiple sources, cite them confidently, and produce coherent reasoning chains, yet the critical piece of information is simply absent. This distinguishes retrieval failure from shared bias, where evidence is present but misinterpreted, and from hallucination, where agents fabricate information beyond what the evidence contains.

\subsubsection{Temporal Reasoning Error}

A temporal reasoning error occurs when agents misinterpret the relationship between evidence timestamps and the resolution window, treating the absence of contemporaneous reporting as a negative signal rather than recognizing an evidence gap.

A representative example is the question (Row 0726) ``If the value of the S\&P 500 index value starting Mar 19, 2025 and ending on Jan 1, 2026 is above 6699.99, then the market resolves to Yes'' (ground truth: YES). The evidence packet contained ten sources, all forecasts or projections published in 2023--2024, with none providing actual observed index values from the resolution window. All three agents converged on NO, reasoning that ``forecasts do not count as evidence of an actual crossing.'' This logic is correct in isolation, but the agents conflated two distinct situations: ``the event has not yet occurred'' and ``the event occurred but our evidence does not include post-event reporting.'' The resolution window had already closed and the S\&P~500 had exceeded the threshold, but the evidence packet contained only pre-event forecasts. Rather than expressing appropriate uncertainty, the agents treated the absence of observed data as affirmative evidence that the threshold was never crossed.

This failure mode is particularly insidious because each individual reasoning step is defensible. The error lies in the implicit temporal assumption that if an event had occurred, the evidence packet would contain confirmation. This assumption is violated whenever the retrieval layer returns temporally misaligned sources, a pattern that recurs across Crypto and Financials questions requiring precise price data at specific timestamps.

\subsubsection{Hallucination}

A hallucination occurs when an agent generates claims or interpretations not supported by the evidence packet or resolution criteria, fabricating content that is presented with the same confidence as legitimate reasoning.

A representative example is the question ``If Star appears on Taylor Swift's album `The Life of a Showgirl' before Jan 1, 2026, then the market resolves to Yes'' (ground truth: YES). The resolution criteria specify rules for matching words in lyrics, establishing that the question asks whether the word ``Star'' appears in the album's lyrics. The evidence packet contained multiple sources with titles such as ``Lyrics Revealed'' and ``Lyrics and Track Listing,'' providing extensive lyrical coverage. However, one model reframed the question entirely, interpreting the resolution criterion as asking whether ``a person/entity named `Star' appears on Taylor Swift's album'' as a performer or contributor. From this fabricated premise, the model concluded NO with 0.75 confidence. The other two models similarly concluded NO, reporting that the word ``Star'' was not mentioned across the lyric sources despite the evidence packet containing articles that comprehensively covered the album's lyrics.

This example illustrates two distinct hallucination patterns. The first is criteria fabrication, where the model invents an interpretation of the resolution criteria that diverges from the actual text. The second is negative assertion from incomplete extraction, where models claim a word does not appear in sources whose full content they may not have fully processed. Both patterns produce reasoning grounded in fabricated premises rather than actual evidence.


\subsubsection{Shared Bias}

A shared bias occurs when all three agents independently arrive at the same incorrect answer due to a common interpretive tendency, despite the relevant evidence being present in the packet. Unlike persuasive error propagation, no inter-agent interaction is involved. This failure mode is the direct cause of the correlated errors measured in Section~\ref{sec:errcorr} and explains why majority voting cannot correct a substantial fraction of errors.

A representative example is the question ``If the first budget reconciliation bill to become law before Jan 1, 2027 cuts at least \$800 billion from Medicaid, then the market resolves to Yes'' (ground truth: YES). The evidence packet contained sources estimating approximately \$625 billion in Medicaid cuts under the House bill and indicating the Senate bill would cut over \$200 billion more, implying roughly \$825 billion in total reductions. All three models correctly identified these figures and recognized that the Senate bill would likely exceed the threshold. Yet all three independently concluded NO, applying identical reasoning: because no source explicitly confirmed that a reconciliation bill had been signed into law, the resolution criterion requiring a bill to ``become law'' was not satisfied. Each model treated the absence of explicit enactment confirmation as decisive, despite the evidence strongly suggesting the threshold would be met.

Because this conservative disposition is shared across model families, neither majority voting nor confidence weighting can surface a correct minority opinion. The error is invisible to the aggregation layer precisely because every agent agrees.

\subsubsection{Persuasive Error Propagation}
\label{sec:failmode-pep}

The four failure modes above apply to both architectures. Persuasive error propagation is unique to Architecture~B and represents the mechanism by which deliberation degrades rather than improves accuracy.

\begin{table}[t]
\centering
\caption{Accuracy by convergence stage in Architecture B's deliberative process. Round~1 consensus = all three models agreed after initial independent answers. Round~2 consensus = models converged after viewing each other's reasoning. No consensus = models still disagreed after both rounds. $n$ = number of questions reaching each stage.}
\label{tab:convergence}
\begin{tabular}{lrc}
\toprule
\textbf{Stage} & $n$ & \textbf{Accuracy (\%)} \\
\midrule
Round 1 consensus     & 1,002 & 80.24 \\
Round 2 consensus     &    58 & 56.90 \\
No consensus reached  &   129 & 52.71 \\
\bottomrule
\end{tabular}
\end{table}

Table~\ref{tab:convergence} shows accuracy stratified by when consensus was reached during Architecture~B's deliberative process. The vast majority of questions (84.3\%, $n = 1{,}002$) reached consensus after Round~1 with 80.24\% accuracy, close to the single-model baseline range, since initial agreement essentially reflects that all three models independently arrived at the same answer. The remaining 187 questions (15.7\%) are substantially harder, with accuracy near chance regardless of whether Round~2 deliberation achieved consensus (56.9\%) or not (52.7\%). The marginal difference between these two groups is 4.2 percentage points, which indicates that when models initially disagree on genuinely difficult questions, further debate rarely resolves the disagreement correctly.

\begin{figure}[t]
\centering
\includegraphics[width=0.75\textwidth]{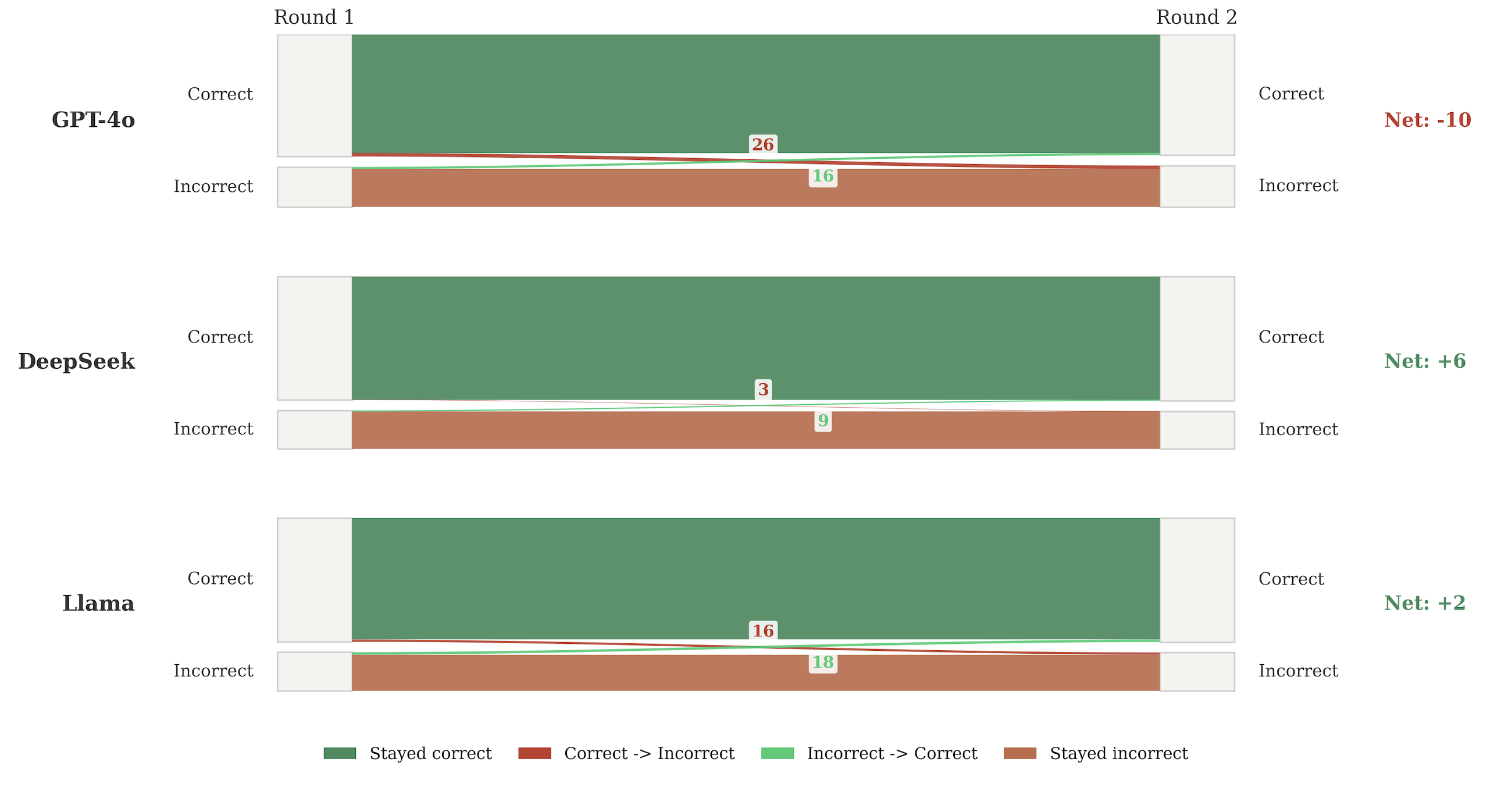}
\caption{\textbf{Answer revision flows between Round~1 and Round~2 in Architecture~B.} Red indicates correct-to-incorrect flips, and green indicates incorrect-to-correct flips. Deliberation yields a net loss in accuracy for GPT-4o (--10) and gains for DeepSeek (+6).}
\label{fig:revision_flows}
\end{figure}

To directly measure the effect of deliberation on individual agent decisions, we tracked all revisions between Round~1 and Round~2 of Architecture~B. Of 3{,}567 agent-round pairs, only 88 (2.47\%) resulted in a revision, affecting 83 of 1{,}189 questions (6.98\%). Critically, these revisions were no better than random: 45 flipped a correct answer to incorrect, while 43 flipped an incorrect answer to correct, yielding a net loss of 2 correct predictions. Per-model analysis reveals an asymmetry consistent with the error correlation findings from Section~\ref{sec:errcorr}. GPT-4o, the most independently-minded model with the lowest pairwise error correlations, was the most harmed by deliberation, flipping 26 correct answers to incorrect against only 16 in the opposite direction. DeepSeek, which shares the highest error correlation with Llama ($r = 0.689$), was the only model to benefit from revisions (3 correct-to-incorrect versus 9 incorrect-to-correct). This pattern suggests that the most independent model is disproportionately vulnerable to being talked out of correct minority positions by a correlated majority, precisely the dynamic that undermines the theoretical promise of multi-agent deliberation. Figure~\ref{fig:revision_flows} visualizes these revision flows for each model.

The mechanism behind this failure appears to be what we term \emph{persuasive error propagation}. During deliberation, a model that is confidently wrong can present compelling but flawed reasoning that causes a correct model to revise its answer. Because the questions where models initially disagree are precisely the ambiguous ones where evidence supports multiple interpretations, a well-articulated wrong argument is often indistinguishable from a right one. This dynamic is the opposite of what Du et al.~\cite{du2023debate} observed in arithmetic tasks, where correct reasoning is verifiable step-by-step and errors are easy to identify during debate. Oracle resolution lacks this verifiability: the correct interpretation of ambiguous evidence cannot be derived through logic alone, making deliberation vulnerable to confident miscalibration.

This finding carries practical design implications. For oracle systems where fast, reliable resolution is the priority, independent aggregation with confidence weighting is both simpler and more accurate than deliberation. Deliberation's computational cost, which requires multiple rounds of inference with full context windows containing other models' reasoning, is not justified by improved accuracy. If anything, the additional expense buys worse outcomes. The one exception, as noted in Section~\ref{sec:category}, is that deliberation occasionally helps on information-dense categories like Companies and World events, where synthesizing multiple interpretive perspectives may add genuine value. A potential hybrid design could route questions through independent aggregation by default, reserving deliberation only for categories where it has demonstrated benefit. This would require sufficient category-level data to calibrate routing decisions reliably.

\subsection{Escalation Criteria for Hybrid Oracle Systems}
\label{sec:escalation}
 
In a deployed oracle system, auto-resolving all questions with Architecture~A yields 82.93\% accuracy. While this represents a meaningful improvement over single-model baselines, the remaining 17\% error rate is substantial for a system settling financial contracts. Human arbitration can achieve higher accuracy but at significant cost and latency. UMA's dispute resolution mechanism requires 48 to 96 hours and involves token-holder voting, while Kalshi's internal operations team provides reliable resolution but is extremely time intensive when scaled to thousands of simultaneous markets. A practical hybrid system needs principled routing rules that auto-resolve high-confidence questions while escalating uncertain ones to human review.
 
We find that inter-agent agreement is the single strongest predictor of resolution accuracy in Architecture~A. Table~\ref{tab:agreement_status} shows accuracy stratified by agreement status across all 1,189 questions. When all three models independently produce the same answer, accuracy reaches 88.34\%. When the vote splits 2-1, accuracy drops to 57.82\%, barely above chance for a binary task. This 30.5 percentage point gap demonstrates that disagreement among independently reasoning models is a highly informative signal that a question is likely to be resolved incorrectly.
 
\begin{table}[h]
\centering
\caption{Architecture~A resolution accuracy by inter-agent agreement status across all 1,189 questions. Unanimous indicates all three models produced the same binary decision; Split indicates a 2-1 vote.}
\label{tab:agreement_status}
\begin{tabular}{lcc}
\hline
Agreement Status & $n$ & Accuracy (\%) \\
\hline
Unanimous (3-0) & 978 & 88.34 \\
Split (2-1) & 211 & 57.82 \\
\hline
\end{tabular}
\end{table}
 
Average model confidence provides additional discriminative power, particularly within the unanimous group. Table~\ref{tab:signal_2x2} combines agreement status with confidence, splitting questions at the median average confidence of 0.91. The unanimous and high-confidence cell achieves 97.87\% accuracy on 563 questions, representing the system's most reliable operating regime. Within unanimous questions, confidence separates near-perfect resolution (97.87\%) from substantially weaker performance (75.42\%). However, for split questions, confidence provides almost no additional signal: accuracy is 56.52\% for high-confidence splits and 58.18\% for low-confidence splits. This asymmetry indicates that agreement is the primary escalation signal, with confidence serving as a useful refinement only when models agree.
 
\begin{table}[h]
\centering
\caption{Architecture~A resolution accuracy by combined agreement status and average confidence. Confidence is split at the median average confidence score (0.91) across all 1,189 questions.}
\label{tab:signal_2x2}
\begin{tabular}{lcccc}
\hline
 & \multicolumn{2}{c}{High Confidence} & \multicolumn{2}{c}{Low Confidence} \\
 & $n$ & Accuracy (\%) & $n$ & Accuracy (\%) \\
\hline
Unanimous & 563 & 97.87 & 415 & 75.42 \\
Split & 46 & 56.52 & 165 & 58.18 \\
\hline
\end{tabular}
\end{table}
 
\begin{figure}[h]
\centering
\includegraphics[width=\textwidth]{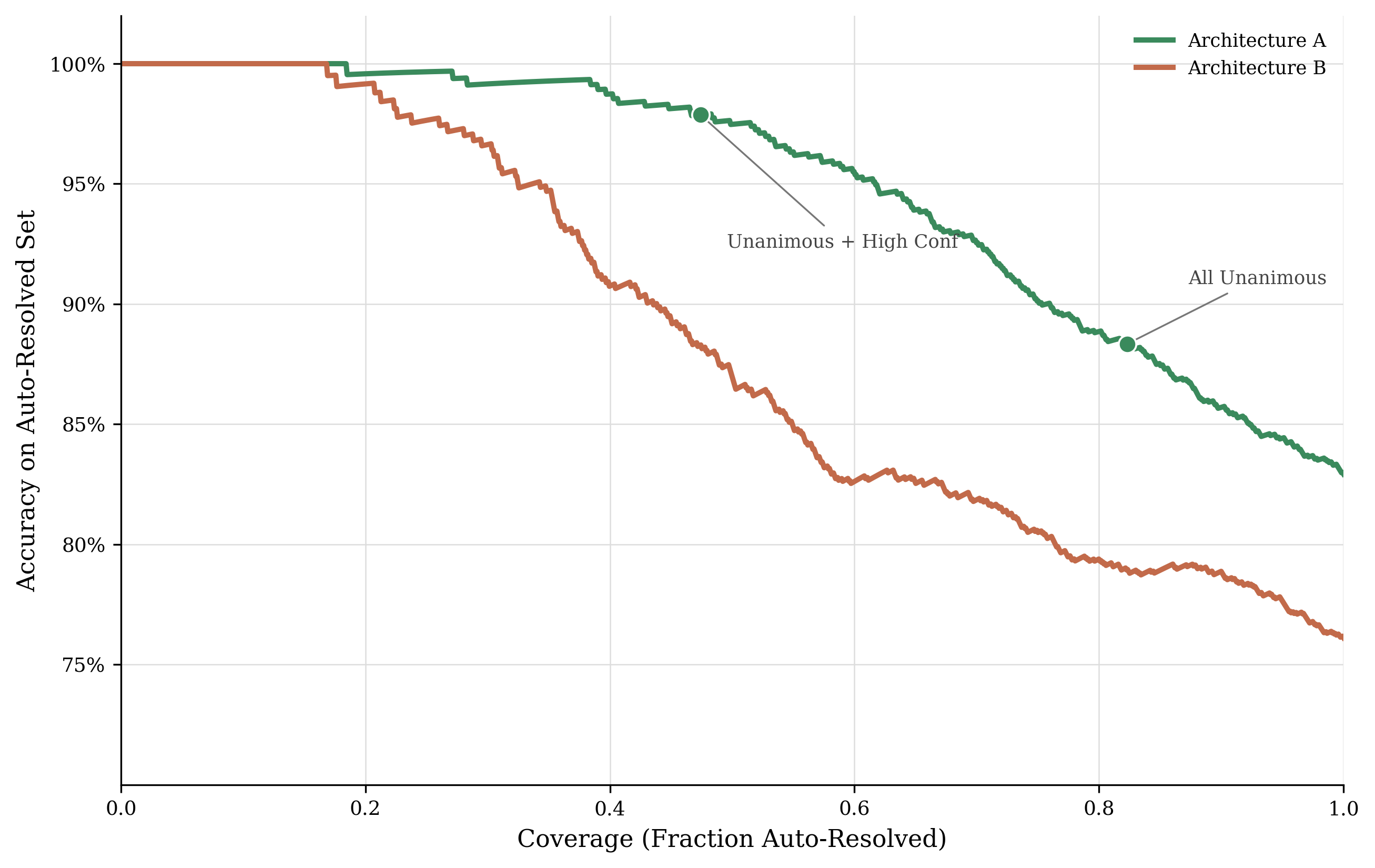}
\caption{\textbf{Coverage-accuracy tradeoff curves for Architecture~A and Architecture~B}. Each point represents the cumulative accuracy when auto-resolving the top $k$ questions ranked by composite escalation score. Architecture~A dominates across most of the operating range.}
\label{fig:coverage_accuracy}
\end{figure}

Figure~\ref{fig:coverage_accuracy} presents the coverage-accuracy tradeoff curve for both architectures, constructed by ranking questions in descending order of composite escalation score and computing cumulative accuracy as coverage expands. Architecture~A achieves 100\% accuracy when auto-resolving only the top 10\% of questions, 97.47\% at 50\% coverage, and 90.12\% at 75\% coverage. Architecture~A dominates Architecture~B across most of the operating range, with the gap reaching 10.44 percentage points at 50\% coverage. Architecture~B is marginally better at the extreme top of the ranking (100\% vs.\ 99.15\% at 10\% coverage), but this advantage disappears by 20\% coverage and Architecture~A is consistently superior thereafter.
 
These results suggest a practical escalation policy for deployed oracle systems. If a system auto-resolves only unanimous, high-confidence questions, it covers 47.4\% of the dataset at 97.87\% accuracy, escalating the remaining 52.6\% to human arbitration. Relaxing the threshold to include all unanimous questions regardless of confidence increases coverage to 82.3\% at 88.34\% accuracy. An oracle designer can select a point on the coverage-accuracy curve that matches their accuracy tolerance and arbitration budget. For context, at 50\% coverage Architecture~A approaches the accuracy levels that Chainlink reported for sports questions (99.7\%) while handling a much broader range of question categories.
 
This analysis treats escalation as a post-hoc empirical investigation rather than a trained decision system. Formal learning-to-defer approaches jointly optimize a classifier and a rejection function to maximize system-level performance \cite{mozannar2020consistent}. Recent work \cite{jung2025trust} has extended this to cascaded LLM systems with provable guarantees of human agreement at user-specified risk tolerances. Applying these formal frameworks to oracle resolution, where the deferral target is human arbitration with known cost and accuracy characteristics, represents a natural direction for future work.

\section{Related Work}

The intersection of large language models, multi-agent systems, and blockchain oracle design has seen rapid development, but no prior work has systematically evaluated multi-LLM oracle architectures for prediction market resolution. This section surveys the most relevant research threads and identifies gaps in existing literature.

\subsection{LLM-Based Oracle Systems}

The most directly relevant prior work is Chainlink Labs' 2025 study evaluating a single-LLM oracle built with the DSPy framework~\cite{chainlink2025aioracles}. Their system uses GPT-4o across three modules: a Question Transformer employing chain-of-thought reasoning, an Information Gathering module performing web scraping with full webpage content retrieval, and a Fact Extraction module producing explicit reasoning traces. Evaluated on 1,660 resolved Polymarket outcomes each with over \$100,000 in trading volume, the system achieved 89.3\% overall accuracy, with dramatic category-dependent variation: 99.7\% on sports, 85.0\% on cryptocurrency, and 84.3\% on political outcomes. The primary bottleneck identified was \emph{temporal cognition}: LLMs struggle with relative time expressions and maintaining awareness of when information was created relative to events being verified. Sports questions, with their discrete temporal endpoints, dramatically outperform politics and cryptocurrency, where temporal boundaries are fuzzy and developments are ongoing. Notably, the study did not filter for potential training data contamination, leaving open the possibility that some accuracy reflects memorized rather than retrieved information. Additionally, their results are also possible skewed given the fact that they only resolved markets with high trading volumes, meaning there are a plethora of information about these markets

UMA's OOTruthBot represents a deployed AI oracle integration that evolved from a single Perplexity API call to a three-layer architecture~\cite{uma2025truthbot}. A lightweight Router dispatches questions to specialized solvers: a Code-Runner for deterministic API queries and a Perplexity-based solver for open-ended questions. An Overseer cross-references outputs against Polymarket live prices as a quality filter. The system achieves approximately 78\% overall accuracy (up to 95\% on objective events with clear data) at roughly \$0.005 per resolution, orders of magnitude cheaper than human-mediated resolution. A notable safety feature is the ``Too Early'' output when evidence is insufficient, implementing a principled abstention mechanism that prevents the system from guessing when it lacks confidence.

\subsection{Multi-Agent Architectures for Improved Reasoning}

The use of multiple LLM instances to improve accuracy has been explored through
several architectural paradigms that differ in how agents interact: independently
(ensemble voting), iteratively (debate and consensus), or hierarchically (layered
aggregation).

The foundational work on multi-agent debate by Du et al.\ demonstrated that
structured argumentation between LLM instances improves factual accuracy and
reasoning~\cite{du2023debate}. Using three agents over two rounds of debate, they
observed arithmetic accuracy improvements from approximately 70\% to 95\%. Two
findings are particularly relevant to oracle design. First, providing full reasoning
chains during debate substantially outperforms sharing only final answers, as agents
benefit from understanding \emph{why} other agents reached their conclusions. Second,
cross-model debate is effective: ChatGPT and Bard, each producing incorrect
individual answers, reached correct solutions through debate. Both observations
motivate our deliberative consensus architecture, which shares full reasoning traces
across diverse model families.

Omar et al.\ introduced the Iterative Consensus Ensemble (ICE), where three LLMs
critique each other iteratively until convergence~\cite{omar2025ice}. ICE improved
overall accuracy from 60.2\% to 74.0\% across over 4,000 questions, with most
questions settling within 2--3 rounds. This convergence behavior suggests that
extended deliberation yields diminishing returns, an observation relevant to the
cost-accuracy tradeoffs central to oracle system design.

\subsection{When Multi-Agent Approaches Fail}
The most important counterpoints to multi-agent enthusiasm come from recent empirical investigations of failure modes. Recent work formalized a ``tyranny of the majority'' effect: as the number of agents providing the same answer increases, whether correct or not, minority agents conform, creating echo chambers rather than enabling error correction~\cite{estornell2024tyranny}. Three interventions were proposed: diversity-pruning, quality-pruning, and misconception-refutation. Complementary work found that group accuracy frequently degrades over debate rounds, and that even introducing a weaker model into a majority of stronger models can diminish overall performance~\cite{talkisntcheap2025}. A large-scale trace analysis examined over 1,600 annotated traces across seven multi-agent frameworks, identifying 14 unique failure modes in three categories, with inter-agent misalignment as the dominant failure type~\cite{cemri2025multiagent}. Using 16$\times$ more compute reduced failure rates by only 43\%, while hierarchical structures outperformed flat architectures by 10--15\%.

The connection to group decision-making theory is instructive. A foundational distinction separates ``demonstrable'' tasks, where a correct answer can be logically proven, from ``judgmental'' tasks where no objectively verifiable solution exists~\cite{laughlin2011group}. For demonstrable tasks, a ``Truth Wins'' dynamic allows even a single correct minority member to persuade the group; for judgmental tasks, a ``Majority Wins'' dynamic dominates regardless of correctness. Oracle resolution straddles both categories: sports outcomes with official results resemble demonstrable tasks, while politically ambiguous questions resemble judgmental ones where convergence may reflect shared biases rather than accuracy.

\subsection{Correlated Errors and the Limits of Aggregation}

Ensemble methods rely on error independence. The Condorcet Jury Theorem guarantees that majority voting improves accuracy only when individual errors are uncorrelated. However, Kim et al.\ find that LLMs frequently agree on the same incorrect answer, indicating substantial error correlation across models~\cite{kim2025correlated}. Related work documents homogenization effects introduced by alignment training~\cite{wu2024monoculture, bai2025systematic}.

These results imply that naive majority voting may yield limited robustness gains. A key question is whether structured deliberation can reduce correlated error or whether it amplifies shared blind spots.

\subsection{Hybrid AI-Human Systems and Principled Escalation}

The question of when an AI system should defer to human judgment is formalized in the learning-to-defer literature, which provides the theoretical grounding for our escalation framework.

Mozannar and Sontag provided a consistent surrogate loss that jointly learns a classifier and rejector~\cite{mozannar2020consistent}. The Bayes-optimal deferral rule has an intuitive form: defer when the probability that the expert is correct exceeds the model's maximum posterior probability for any class. In oracle terms, this means deferring to human arbitration when the AI system's confidence is lower than expected human accuracy, a principled cost-benefit calculation rather than an ad-hoc threshold.

Bansal et al.\ found that AI explanations did not increase human-AI team accuracy beyond simply showing confidence scores, as explanations increased acceptance of AI recommendations regardless of correctness~\cite{bansal2021does}. This finding cautions against assuming that providing LLM reasoning traces to human arbitrators in an escalation pipeline will improve their judgment; confidence signals and disagreement patterns may be more actionable.

Selective prediction formalizes the tradeoff between accuracy and coverage: a system that declines to answer uncertain questions will be more accurate on those it does answer, but at the cost of resolving fewer questions automatically~\cite{chow1970optimum, geifman2017selective}. UMA's OOTruthBot already implements principled abstention through its ``Too Early'' output~\cite{uma2025truthbot}. Recent work formalizes three-tier LLM-human architectures where a base model routes to a large model which routes to a human expert~\cite{tiered2025architecture}, a pattern that maps directly to oracle system design. Our escalation framework builds on these approaches by using multi-agent disagreement and confidence as principled routing signals.

\subsection{Gaps in Existing Research}

Across these threads, no work has systematically evaluated multi-LLM
architectures for prediction market resolution, characterized error
correlation in oracle settings specifically, or developed principled
escalation criteria grounded in the learning-to-defer framework. The
methodology in Section~\ref{methodology} addresses these gaps through a controlled
comparison of independent aggregation and deliberative consensus, an
error correlation analysis across model pairs, and an escalation framework
that uses inter-agent disagreement and confidence as routing signals.

\section{Conclusion}
This paper evaluated whether multi-agent LLM architectures can 
improve oracle resolution accuracy over single-model baselines. 
Using 1,189 resolved prediction market questions from KalshiBench, 
we compared two architectures, independent aggregation and 
deliberative consensus, against single-model baselines in a domain 
where correctness has direct financial consequences.

This work makes three contributions to the intersection of multi-agent 
AI systems and decentralized oracle design: (1) empirical evidence 
that independent aggregation offers modest but consistent accuracy 
gains while deliberation degrades performance in evidence-based 
resolution; (2) a characterization of error correlation patterns 
specific to oracle resolution, showing that shared training data and 
architectural lineage produce failures that fundamentally limit 
aggregation benefits; and (3) escalation criteria that use inter-agent 
disagreement and confidence as principled routing signals for hybrid 
AI-human systems. The remainder of this chapter develops each in turn.

The results yield a nuanced answer to the central research question. 
Independent aggregation with confidence-weighted voting achieves the 
highest accuracy at 83.43\%, a statistically significant improvement 
over the best single-model baseline, with the largest gains in 
interpretively ambiguous categories like Politics (+3.2 percentage 
points). Yet these gains fall well short of the theoretical Condorcet 
ceiling of approximately 90\%, a gap explained by moderate-to-high 
error correlations across model pairs ($r = 0.529$--$0.689$): when 
one model fails, the others typically fail on the same question for 
the same underlying reason. The same bound shows up at the category 
level. Well-documented domains like Sports (92.9\%) and Financials 
(93.3\%) approach the ceiling reported by prior single-LLM studies, 
while ambiguous domains like Crypto (66.1\%) and Companies (58.0\%) 
resist improvement from any architecture. Together, these findings 
leave a 14\% hard core of questions that no multi-agent approach 
resolves, motivating principled escalation to human arbitration.

The most striking finding is that deliberative consensus degrades performance rather than improving it. All final-round outputs in Architecture B fall below every single-model baseline, with the best deliberative result trailing the worst individual model by over four percentage points. We attribute this to persuasive error propagation and LLM sycophancy: on genuinely difficult questions where models initially disagree, a confidently wrong model can present compelling but flawed reasoning that causes a correct model to revise its answer. This stands in contrast to prior work on multi-agent debate for arithmetic and logical reasoning tasks, where correct derivations are verifiable step-by-step. Oracle resolution lacks this verifiability, making deliberation vulnerable to confident miscalibration.

These findings carry practical implications for the design of oracle systems as prediction markets continue to grow in volume and complexity. For deployed systems where fast, reliable resolution is the priority, independent aggregation with confidence weighting is both simpler and more accurate than deliberation. The escalation framework provides a structured alternative to ad-hoc human override policies: unanimous, high-confidence resolutions achieve 97.87\% accuracy while covering nearly half of all questions, enabling oracle systems to auto-resolve the easy cases while routing genuinely uncertain questions to human arbitration.

More broadly, these findings suggest that the benefits of multi-agent LLM systems may be fundamentally constrained by correlated reasoning errors across models. In domains where correctness depends on interpreting ambiguous real-world evidence rather than deriving verifiable logical conclusions, debate-based architectures risk amplifying persuasive but incorrect arguments rather than correcting them. As decision-making systems increasingly operate as human-AI hybrids, designing reliable AI oracle systems may require architectures that prioritize independence and calibrated escalation over deliberative consensus.

\section{Future Work}
\subsection{Pipeline Architecture Improvements}

\begin{figure}[t]
    \centering
    \includegraphics[width=\linewidth]{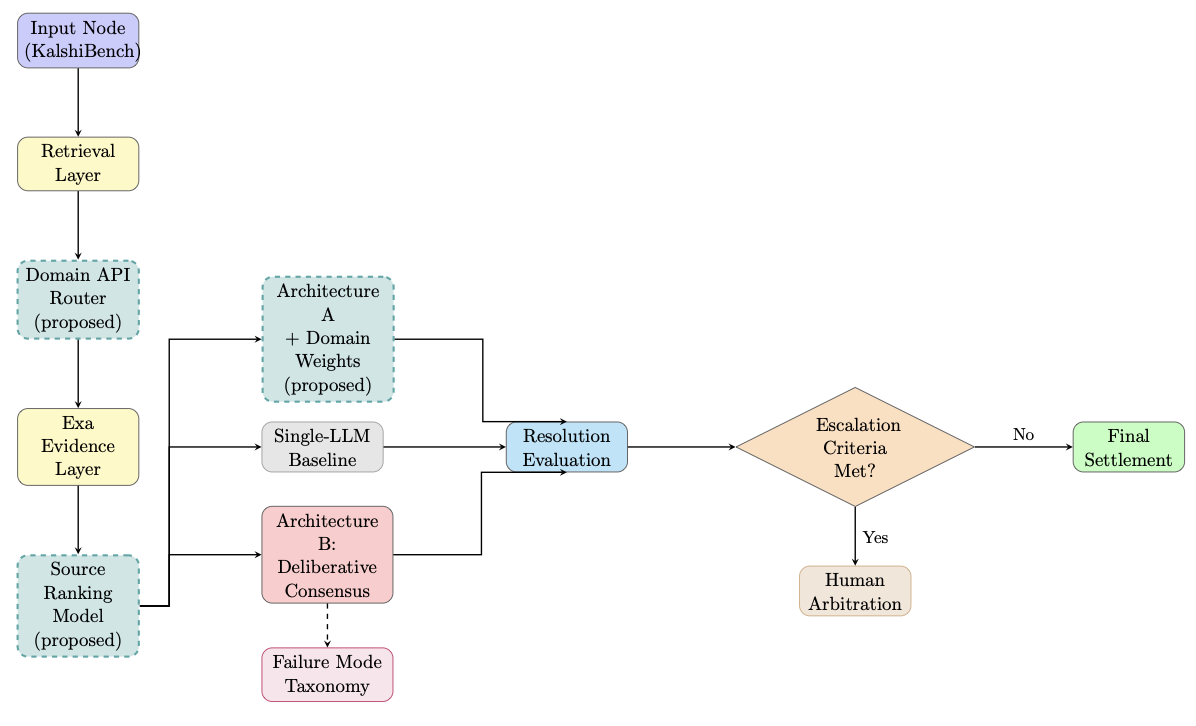}
    \caption{\textbf{Extended system flow with proposed future-work updates.} 
    The pipeline introduces domain-specific retrieval routing, learned source ranking, and domain-aware aggregation to improve retrieval quality, evidence prioritization, and ensemble performance.}
    \label{fig:future_system}
\end{figure}

We identify several complementary directions for improving the oracle pipeline that extend its retrieval, evidence selection, and aggregation layers. Figure~\ref{fig:future_system} illustrates the proposed modifications.

\subsubsection{Specialized Retrieval for Domain-Specific Questions}

A recurring source of error in our evaluation was questions requiring precise, structured data that general-purpose search engines are poorly suited to retrieve. For example, a question such as "Will the price of Bitcoin be above \$80,000 on March 24th, 2025?" requires a single authoritative price observation at a specific timestamp. This information is trivially available through a cryptocurrency price API like CoinGecko but difficult to extract reliably from web search results, which may return commentary, forecasts, or prices from neighboring dates. Our use of Exa as a uniform retrieval backend for all question types meant that such domain-specific queries often received evidence packets that were insufficient for confident resolution. A more robust system would combine a general-purpose retrieval layer for open-ended questions with specialized API integrations dispatched based on question category or content. For categories like Crypto or Financials, a routing layer could detect that the question requires a structured data lookup and query the appropriate domain API directly, using cheap and deterministic lookups for questions that warrant them and general search for everything else.

\subsubsection{Source Quality Weighting and Learned Source Ranking}

Our current pipeline treats all retrieved sources equally, passing the same unweighted evidence packet to every agent. However, not all sources are equally informative or reliable for a given question: a primary government data release carries more evidentiary weight than a speculative blog post. A natural extension would be to introduce a learned source-ranking layer that scores and weights retrieved documents before they reach the resolution agents. One promising approach would train an embedding model to reproduce quality rankings generated by LLM judges applied to oracle resolution traces. Specifically, for a set of resolved questions with known ground-truth outcomes, one could use one or more LLM judges to rank the sources in each evidence packet by their contribution to a correct resolution. These ranked preferences would then serve as training signal for a lightweight source-ranking model that could be applied at inference time without the cost of repeated LLM judge calls. Evaluation of such a system is straightforward in principle. By holding the resolution architecture fixed, one can measure whether accuracy improves when agents receive source-weighted evidence compared to the uniform-weight baseline used in this study. Beyond accuracy, source weighting could also improve calibration by allowing agents to express lower confidence when their highest-weighted sources are weak, providing a more informative signal for the escalation framework developed in this thesis.

\subsubsection{Principled Abstention as a Third Resolution Output}

Our current agent schema forces a binary decision. Every agent must output YES or NO for every question, regardless of evidence quality. While agents can express low confidence, they cannot explicitly signal that the evidence is insufficient to resolve the question. This constraint is a direct contributor to several failure modes identified in Section~\ref{sec:failure}. In retrieval failures, agents produce confident but wrong answers because the evidence packet appears substantive even when the critical piece of information is absent. In temporal reasoning errors, agents correctly recognize that their evidence contains only forecasts rather than observed outcomes, but are forced to pick a side rather than flag the evidence gap.

Adding abstention changes the aggregation logic in meaningful ways. Under the current majority-vote scheme, a 2-1 split triggers escalation while a 3-0 unanimous answer proceeds automatically. With abstention, new configurations arise. If two agents abstain and one votes YES, the system should escalate rather than treat a single vote as a resolution. More importantly, abstention could surface failure cases that the current escalation framework cannot detect. The unanimous-but-wrong resolutions in our evaluation, which account for a substantial portion of errors, appear indistinguishable from unanimous correct resolutions under the current routing policy. If agents could abstain when evidence is thin, some of these cases would produce mixed vote-abstain patterns that trigger human review rather than proceeding to automatic settlement.

\subsubsection{Asymmetric Model Weighting and Ensemble Diversity}

The current aggregation mechanism treats all three oracle agents as equals, applying a uniform confidence-weighted vote regardless of domain. However, the category-level results suggest this is suboptimal. In Elections, a single model occasionally held the correct minority position that majority aggregation overrode, diluting a well-calibrated signal with noisier ones. A natural extension within Architecture~A would be to assign per-model influence based on demonstrated category-level accuracy, effectively building a domain-aware ensemble that up-weights the historically strongest model for each question type. Such a weighting scheme could be derived from held-out validation data and applied at inference time without additional LLM calls. A related question is whether expanding the ensemble with a fourth, more architecturally distinct model would meaningfully reduce the error correlations that limit aggregation gains. The moderate-to-high pairwise correlations observed between DeepSeek and Llama ($r = 0.689$) suggest that open-weight models sharing overlapping training lineages inherit shared reasoning blind spots, and introducing a model trained with a fundamentally different optimization objective could widen ensemble diversity and push majority-vote accuracy closer to the Condorcet ceiling that correlated errors currently prevent.

\subsection{Adversarial Robustness and Evidence Injection Attacks}

A limitation not addressed in this study is the robustness of multi-agent oracle architectures to adversarial manipulation of the retrieval layer. If a bad actor plants misleading but plausible sources in the web index, both architectures receive the same poisoned evidence packet, since all agents share a common retrieval layer. Under independent aggregation, correlated errors from a shared adversarial source would propagate directly into the majority vote. Under deliberative consensus, the concern is more subtle. A persuasively written but fabricated source could anchor one agent's reasoning and then influence the others during deliberation, amplifying rather than containing the attack. Future work could evaluate robustness by injecting synthetic adversarial documents into evidence packets at varying rates and measuring accuracy degradation across architectures. A secondary question is whether inter-agent disagreement, which serves as an escalation signal in this thesis, also functions as an adversarial detection signal. A poisoned source that convinces two of three agents but not the third would surface as a split vote and trigger human review under the routing policy proposed in Section~\ref{sec:escalation}, suggesting that the escalation framework may provide partial robustness against certain classes of evidence injection attacks even without explicit defenses.

\subsection{Formal Learning-to-Defer Integration}

The escalation framework developed in Section~\ref{sec:escalation} treats routing as a post-hoc empirical analysis, constructing a composite score from inter-agent agreement and average confidence and evaluating its discriminative power on the same dataset used for resolution. While this approach yields interpretable and practically useful routing thresholds, it does not jointly optimize the resolution and deferral functions. \cite{mozannar2020consistent} provide a consistent surrogate loss that jointly learns a classifier and a rejector, where the Bayes-optimal deferral rule defers to the human expert whenever the probability that the expert is correct exceeds the model's maximum posterior probability for any class. Applying this framework to oracle resolution would mean training a rejector that takes multi-agent outputs as features and learns directly from resolution outcomes and human arbitration accuracy, rather than relying on a hand-crafted composite score. A natural evaluation would compare the coverage-accuracy tradeoff curve achieved by a trained rejector against the empirical curve reported in Figure~\ref{fig:coverage_accuracy}, measuring whether principled joint optimization meaningfully outperforms the agreement-plus-confidence heuristic. This direction is particularly promising for oracle system design because the deferral target, human arbitration via mechanisms like UMA's Data Verification Mechanism, has known and measurable accuracy and cost characteristics, satisfying the conditions under which the Bayes-optimal deferral rule yields provable guarantees.

\section*{Acknowledgments}

I would first like to thank Professor Fan Zhang for his immeasurable support and valuable guidance throughout this process. His expertise in oracle design was invaluable to this project.

I would also like to thank the Yale Department of Computer Science for fostering an environment that sparked my curiosity and encouraged a lifelong pursuit of exploration in the field.

Finally, I am deeply grateful to my parents, whose unwavering love and early academic inspiration shaped me into the student I am today, and to my twin sister, for sharing so much of my academic journey with me and for being a damn good friend.

\printbibliography

@article{arrow2008promise,
  title={The promise of prediction markets},
  author={Arrow, Kenneth J and Forsythe, Robert and Gorham, Michael and Hahn, Robert and Hanson, Robin and Ledyard, John O and Levmore, Saul and Litan, Robert and Milgrom, Paul and Nelson, Forrest D and others},
  journal={Science},
  volume={320},
  number={5878},
  pages={877--878},
  year={2008},
  publisher={American Association for the Advancement of Science}
}

@article{berg2008prediction,
  title={Prediction market accuracy in the long run},
  author={Berg, Joyce E and Nelson, Forrest D and Rietz, Thomas A},
  journal={International Journal of Forecasting},
  volume={24},
  number={2},
  pages={283--298},
  year={2008},
  publisher={Elsevier}
}

@misc{kalshi2024cftc,
  title={US election betting: {CFTC} loses last-minute bid to halt {Kalshi} contract},
  author={{CoinDesk}},
  year={2024},
  howpublished={\url{https://www.coindesk.com/policy/2024/09/12/us-election-betting-cftc-loses-last-minute-bid-to-halt-kalshi-contract}}
}

@article{caldarelli2020oracle,
  title={Understanding the blockchain oracle problem: A call for action},
  author={Caldarelli, Giulio},
  journal={Information},
  volume={11},
  number={11},
  pages={509},
  year={2020},
  publisher={MDPI}
}

@misc{chainalysis2023oracle,
  title={Oracle Manipulation Attacks Rising: A Unique Concern for {DeFi}},
  author={{Chainalysis Team}},
  howpublished={Chainalysis Blog},
  year={2023},
  url={https://www.chainalysis.com/blog/oracle-manipulation-attacks-rising/}

}

@techreport{ellis2017chainlink,
  title={Chainlink: A decentralized oracle network},
  author={Ellis, Steve and Juels, Ari and Nazarov, Sergey},
  year={2017},
  institution={Chainlink Labs},
  note={\url{https://research.chain.link/whitepaper-v1.pdf}}
}

@misc{chainlink2021v2,
  title={Chainlink 2.0: Next steps in the evolution of decentralized oracle networks},
  author={Breidenbach, Lorenz and Cachin, Christian and Chan, Benedict and Coventry, Alex and Ellis, Steve and Juels, Ari and Koushanfar, Farinaz and Miller, Andrew and Magauran, Brendan and Moroz, Daniel and others},
  year={2021},
  howpublished={\url{https://research.chain.link/whitepaper-v2.pdf}}
}

@misc{uma2024oracle,
  title={How does {UMA}'s oracle work?},
  author={{UMA Protocol}},
  year={2024},
  howpublished={\url{https://docs.uma.xyz/protocol-overview/how-does-umas-oracle-work}}
}

@misc{chainlink2025aioracles,
  title={Empirical evidence in {AI} oracle development},
  author={Zintus-art, Kaspars and Vass, Brandon and Ward, Jonathan},
  year={2025},
  howpublished={\url{https://blog.chain.link/ai-oracles/}}
}

@misc{uma2025truthbot,
  title={Inside {UMA}'s optimistic truth bot},
  author={{UMA Protocol}},
  year={2025},
  howpublished={\url{https://blog.uma.xyz/articles/inside-umas-optimistic-truth-bot}}
}

@article{ji2023hallucination,
  title={Survey of hallucination in natural language generation},
  author={Ji, Ziwei and Lee, Nayeon and Frieske, Rita and Yu, Tiezheng and Su, Dan and Xu, Yan and Ishii, Etsuko and Bang, Ye Jin and Madotto, Andrea and Fung, Pascale},
  journal={ACM Computing Surveys},
  volume={55},
  number={12},
  pages={1--38},
  year={2023},
  publisher={ACM}
}

@article{huang2025hallucination,
  title={A survey on hallucination in large language models: Principles, taxonomy, challenges, and open questions},
  author={Huang, Lei and Yu, Weijiang and Ma, Weitao and Zhong, Weihong and Feng, Zhangyin and Wang, Haotian and Chen, Qianglong and Peng, Weihua and Feng, Xiaocheng and Qin, Bing and Liu, Ting},
  journal={ACM Transactions on Information Systems},
  volume={43},
  number={2},
  pages={1--44},
  year={2025},
  publisher={ACM}
}

@inproceedings{fanous2025sycophancy,
  title={Sycophancy in {LLM}s: Causes, consequences, and mitigation strategies},
  author={Fanous, Amir and others},
  booktitle={Proceedings of the AAAI/ACM Conference on AI, Ethics, and Society},
  volume={8},
  number={1},
  pages={893--900},
  year={2025}
}

@inproceedings{du2023debate,
  title={Improving factuality and reasoning in language models through multiagent debate},
  author={Du, Yilun and Li, Shuang and Torralba, Antonio and Tenenbaum, Joshua B and Mordatch, Igor},
  booktitle={Proceedings of the 41st International Conference on Machine Learning (ICML)},
  year={2024},
  note={arXiv preprint arXiv:2305.14325}
}

@article{omar2025ice,
  title={Refining {LLM} outputs with iterative consensus ensemble ({ICE})},
  author={Omar, Mohamed and Glicksberg, Benjamin S and Nadkarni, Girish N},
  journal={Computers in Biology and Medicine},
  volume={196},
  year={2025},
  publisher={Elsevier}
}

@article{kim2025correlated,
  title={Correlated errors in large language models},
  author={Kim, Eunsu and Garg, Nikhil and Peng, Kaiwen and Garg, Siddharth},
  journal={arXiv preprint arXiv:2506.07962},
  year={2025},
  note={ICML 2025}
}

@inproceedings{estornell2024tyranny,
  title={On the {Tyranny} of the {Majority}: How multi-agent debate can improve upon majority voting},
  author={Estornell, Andrew and Liu, Yang},
  booktitle={Advances in Neural Information Processing Systems (NeurIPS)},
  year={2024}
}

@article{mozannar2020consistent,
  title={Consistent Estimators for Learning to Defer to an Expert},
  author={Mozannar, Hussein and Sontag, David},
  booktitle={Proceedings of the 37th International Conference on Machine Learning (ICML)},
  volume={119},
  pages={7076--7087},
  year={2020},
  publisher={PMLR}
}

@article{jung2025trust,
  title={Trust or Escalate: {LLM} Judges with Provable Guarantees for Human Agreement},
  author={Jung, Jaehun and others},
  booktitle={Proceedings of the International Conference on Learning Representations (ICLR)},
  year={2025}
}

@article{chow1970optimum,
  title={On optimum recognition error and reject tradeoff},
  author={Chow, Chi-Keung},
  journal={IEEE Transactions on Information Theory},
  volume={16},
  number={1},
  pages={41--46},
  year={1970}
}

@article{cemri2025multiagent,
  title={Multi-agent failure mode analysis},
  author={Cemri, Mert and others},
  journal={arXiv preprint arXiv:2503.13657},
  year={2025},
  note={NeurIPS 2025 Spotlight}
}

@article{nel2025kalshibench,
  title={Do large language models know what they don't know?},
  author={Nel, Ethan},
  journal={arXiv preprint arXiv:2512.16030},
  year={2025},
  note={KalshiBench benchmark}
}

@article{hayek1945use,
  title={The use of knowledge in society},
  author={Hayek, Friedrich A},
  journal={American Economic Review},
  volume={35},
  number={4},
  pages={519--530},
  year={1945}
}

@inproceedings{iceb2020meta,
  title={A meta-analysis of prediction markets accuracy},
  author={{ICEB Conference}},
  booktitle={Proceedings of the International Conference on Electronic Business},
  year={2020}
}

@misc{kalshi_outcomes,
  title={Market outcomes},
  author={{Kalshi}},
  year={2024},
  howpublished={\url{https://help.kalshi.com/markets/markets-101/market-outcomes}}
}

@article{cong2025oracle,
  title={A primer on oracle economics},
  author={Cong, Lin William and Fox, Brett and Li, Yizhou and Zhou, Zhiheng},
  journal={Journal of Corporate Finance},
  year={2025},
  publisher={Elsevier}
}

@misc{polymarket_uma_adapter,
  title={{Polymarket UMA CTF} adapter},
  author={{Polymarket}},
  year={2024},
  howpublished={\url{https://github.com/Polymarket/uma-ctf-adapter}}
}

@article{talkisntcheap2025,
  title={Talk isn't always cheap: Understanding failure modes in multi-agent debate},
  author={{Anonymous}},
  journal={arXiv preprint arXiv:2509.05396},
  year={2025}
}

@book{laughlin2011group,
  title={Group Problem Solving},
  author={Laughlin, Patrick R},
  year={2011},
  publisher={Princeton University Press}
}

@article{wu2024monoculture,
  title={Generative monoculture in large language models},
  author={Wu, Tongshuang and others},
  journal={arXiv preprint},
  year={2024}
}

@article{bai2025systematic,
  title={Measuring and addressing systematic bias in {LLM} decision-making},
  author={Bai, Yuntao and others},
  journal={Proceedings of the National Academy of Sciences},
  volume={122},
  number={9},
  year={2025}
}

@inproceedings{thorne2018fever,
  title={{FEVER}: A large-scale dataset for fact extraction and verification},
  author={Thorne, James and Vlachos, Andreas and Christodoulopoulos, Christos and Mittal, Arpit},
  booktitle={Proceedings of the 2018 Conference of the North American Chapter of the Association for Computational Linguistics (NAACL)},
  year={2018}
}

@inproceedings{bansal2021does,
  title={Does the whole exceed its parts? The effect of {AI} explanations on complementary team performance},
  author={Bansal, Gagan and Wu, Tongshuang and Zhou, Joyce and Fok, Raymond and Nushi, Besmira and Kamar, Ece and Ribeiro, Marco Tulio and Weld, Daniel},
  booktitle={Proceedings of the ACM CHI Conference on Human Factors in Computing Systems},
  year={2021}
}

@inproceedings{geifman2017selective,
  title={Selective classification for deep neural networks},
  author={Geifman, Yonatan and El-Yaniv, Ran},
  booktitle={Advances in Neural Information Processing Systems (NeurIPS)},
  year={2017}
}

@article{tiered2025architecture,
  title={Three-tier {LLM}-human cascaded architectures for scalable decision systems},
  author={{Anonymous}},
  journal={arXiv preprint arXiv:2506.11887},
  year={2025}
}

\appendix
\section{Appendix}

\lstset{
  breaklines=true,
  breakatwhitespace=true,
  basicstyle=\ttfamily\small
}

\subsection{Prompt Templates}

\begin{figure}[H]
\centering
\begin{lstlisting}[style=promptstyle, frame=single]
SYSTEM PROMPT:

You are an expert prediction market resolution agent. Your task is to determine whether a prediction market question should resolve to YES or NO based on the provided evidence.

Instructions:
1. Read the question and resolution criteria carefully.
2. Analyze ALL provided sources for relevant information.
3. Make your decision based on definitive evidence. Prioritize information describing outcomes that have already occurred.
4. If evidence is ambiguous, use your best judgment.
5. Rate your confidence from 0.0 (very uncertain) to 1.0 (absolutely certain).

Output Format:
- decision: YES or NO
- confidence: 0.0 to 1.0
- reasoning: Explanation referencing specific sources

USER MESSAGE TEMPLATE:

Please analyze the following prediction market question and evidence, then provide your resolution decision.

{evidence_text}

Based on the evidence above, should this question resolve to YES or NO?
\end{lstlisting}
\caption{\textbf{System prompt used for single-LLM oracle resolution in Architecture A}. The model receives structured evidence and is required to output a binary decision, calibrated confidence score, and source-grounded reasoning.}
\label{fig:system-prompt}
\end{figure}

\begin{figure}[t]
\centering

\begin{tcolorbox}[
    colback=white,
    colframe=black,
    boxrule=0.5pt,
    arc=0pt,
    outer arc=0pt,
    left=4pt,
    right=4pt,
    top=4pt,
    bottom=4pt
]
\begin{lstlisting}[
    style=promptstyle,
    basicstyle=\ttfamily\scriptsize
]
ROUND_1_SYSTEM_PROMPT:
You are an expert prediction market resolution agent. Your task is to determine whether a prediction market question should resolve to YES or NO based on the provided evidence.

Determine whether the question should resolve to YES or NO based only on the evidence provided.

ROUND_1_USER_TEMPLATE:
QUESTION: {question}

RESOLUTION CRITERIA: {criteria}

EVIDENCE PROVIDED from reputable sources:
{exa_evidence}

Provide your answer in JSON with these fields:
- decision: YES or NO
- confidence: number between 0.0 and 1.0
- reasoning: detailed step-by-step reasoning that cites relevant evidence and criteria interpretation

Think carefully about:
1. Read the question and resolution criteria carefully
2. Analyze ALL provided sources for relevant information
3. Make your decision based on definitive evidence. You should prioritize information that describes an outcome that happened.
4. If evidence is ambiguous, use your best judgment
5. Rate your confidence in your decision from 0.0 (very uncertain) to 1.0 (absolutely certain)
\end{lstlisting}
\end{tcolorbox}

\caption{\textbf{Architecture B (Deliberative Consensus), Round 1 prompt}. Each agent independently resolves the market using a shared evidence packet and outputs a structured JSON decision, confidence score, and evidence-grounded reasoning.}
\label{fig:archB-round1-prompt}
\end{figure}

\refstepcounter{figure}\label{fig:archB-round2-prompt}

\noindent
\begin{tcolorbox}[
  breakable,
  colback=white,
  colframe=black,
  boxrule=0.5pt,
  arc=0pt,
  outer arc=0pt
]
\begin{lstlisting}[style=promptstyle]
ROUND_2_SYSTEM_PROMPT:
You are in the final cross-examination round of a multi-agent resolution debate.
You will see how other agents reasoned about the same question. Your job is to CRITICALLY EVALUATE their reasoning against the evidence -- not to defer to them.

IMPORTANT: You should only change your answer if you identify a SPECIFIC factual error in your own round 1 reasoning, or if another agent points to SPECIFIC EVIDENCE in the shared packet that you misread or overlooked. Do NOT change your answer simply because other agents disagree with you or sound confident.

ROUND_2_USER_TEMPLATE:
You previously resolved this prediction market question.
Now you will see how two other expert agents reasoned about the same question.

QUESTION: {question}

RESOLUTION CRITERIA: {criteria}

EVIDENCE (SAME SHARED PACKET FROM ROUND 1):
{exa_evidence}

YOUR PREVIOUS ANSWER:
{your_round_1_response}

OTHER AGENTS' REASONING:
{other_agents_section}

This is the FINAL round. Critically evaluate the other agents' reasoning:

1. For each agent that DISAGREES with you: Do they cite specific evidence from the shared packet that contradicts your reasoning? Or are they asserting a conclusion without evidentiary support?
2. Re-read the specific pieces of evidence that are most relevant to the disagreement.
3. ONLY change your decision if you can identify a concrete error in your own round 1 analysis, for example, you misread a date, overlooked a source, or misinterpreted the resolution criteria.
4. If agents agree with you, do NOT increase your confidence unless they provide additional evidence-based reasoning you hadn't considered.

DEFAULT BEHAVIOR: Change your decision ONLY if another agent identifies specific evidence
or because there is a concrete flaw in your reasoning, not simply because they reached
a different conclusion.

Provide your FINAL answer in JSON with these fields:
- decision: YES or NO
- confidence: number between 0.0 and 1.0
- reasoning: explain your final reasoning, explicitly addressing each disagreeing agent's key claim
- revised: true if you changed your decision from round 1, else false
- convergence_notes: short note on agreement/disagreement and why
\end{lstlisting}
\end{tcolorbox}

\captionof{figure}{\textbf{Architecture B (Deliberative Consensus), Round 2 prompt}. Each agent performs a final cross-examination using the same evidence packet, explicitly evaluates other agents' claims, and revises its decision only if it identifies a concrete evidence-based error in its prior reasoning.}

\subsection{Command-Line Interface Reference}
\label{sec:cli-reference}

The evaluation pipeline exposes two main entry points, \texttt{scripts/test\_architecture\_a.py} and \texttt{scripts/test\_architecture\_b.py}. Both scripts share a common set of command-line flags for dataset selection, evidence retrieval, model configuration, and output control. Table~\ref{tab:cli-flags} documents all available flags. To reproduce the main experimental results, a typical invocation for Architecture A is:

\begin{verbatim}
python scripts/test_architecture_a.py \
    --questions-csv results/kalshibench_subset_450_seed42.csv \
    --retrieval-mode highlights \
    --num-results 10 \
    --cache-dir cache/evidence \
    -n 1189 \
    -o results/architecture_a_results.csv
\end{verbatim}

\noindent and for Architecture B:

\begin{verbatim}
python scripts/test_architecture_b.py \
    --questions-csv results/kalshibench_subset_450_seed42.csv \
    --retrieval-mode highlights \
    --num-results 10 \
    --cache-dir cache/evidence \
    -n 1189 \
    -o results/architecture_b_results.csv
\end{verbatim}

\noindent The \texttt{--cache-only} flag can be passed on subsequent runs to avoid re-issuing Exa queries for questions whose evidence packets are already cached locally.

\begin{table}[ht]
\centering
\caption{Command-line flags for the Architecture A and Architecture B evaluation scripts.
``A'' and ``B'' in the \textit{Applies To} column indicate which script(s) expose the flag.}
\label{tab:cli-flags}
\renewcommand{\arraystretch}{1.3}
\small
\begin{tabular}{p{4cm} p{1.5cm} p{3.5cm} p{5.5cm}}
\toprule
\textbf{Flag} & \textbf{Applies To} & \textbf{Default} & \textbf{Description} \\
\midrule
\texttt{-n} & A, B & 10 & Number of questions to evaluate \\
\texttt{--questions-csv} & A, B & \texttt{results/kalshi-} \newline \texttt{bench\_subset\_} \newline \texttt{450\_seed42.csv} & Path to the input questions CSV \\
\texttt{--start-row} & A, B & 1 & 1-based row number to begin evaluation \\
\texttt{--cache-start-row} & A, B & None & 1-based row for cache-key lookup, independent of \texttt{--start-row} \\
\texttt{--question-ids} & A, B & None & Specific question IDs to evaluate, bypassing sequential sampling \\
\texttt{--category} & A, B & None & Filter the dataset to a single category, such as \texttt{Politics} or \texttt{Sports} \\
\texttt{--retrieval-mode} & A, B & \texttt{highlights} & Exa retrieval mode: \texttt{full\_text} or \texttt{highlights} \\
\texttt{--num-results} & A, B & 10 & Number of Exa sources retrieved per question \\
\texttt{--fulltext-max-\newline characters} & A, B & 4000 & Character cap per source in full-text mode \\
\texttt{--highlights-\newline max-characters} & A, B & 2000 & Character cap for Exa highlight snippets \\
\texttt{--cache-only} & A, B & false & Use local cache only; do not issue live Exa queries \\
\texttt{--cache-dir} & A, B & \texttt{cache/evidence} & Directory for cached evidence packets \\
\texttt{--use-claude} & A, B & false & Substitute Claude for the DeepSeek agent slot \\
\texttt{--use-gemini} & A, B & false & Substitute Gemini for the Llama agent slot \\
\texttt{-o} / \texttt{--output} & A & auto-generated timestamped path & Output path for per-question results CSV; auto-generates a timestamped filename if omitted \\
\texttt{-o} / \texttt{--output} & B & \texttt{arch\_b\_} \newline \texttt{results.csv} & Output path for the per-question results CSV \\
\texttt{--json} & A, B & None & Optional path for aggregate summary JSON \\
\texttt{--dry-run} & A, B & false & Print the planned configuration without executing API calls \\
\texttt{-v} / \texttt{--verbose} & A, B & false & Enable per-question progress logging \\
\bottomrule
\end{tabular}
\end{table}

\subsection{Code Availability}
\label{sec:code-availability}

The full implementation of the multi-agent oracle system, including evaluation scripts, retrieval pipeline, and experiment configurations, is publicly available on GitHub:

\begin{center}
\href{https://github.com/Tkota10/Senior-Thesis-Multi-LLM-Resolution}{\underline{\textcolor{blue!70!black}{Paper Repository}}}
\end{center}

\end{document}